\documentclass{article}
  
\usepackage{graphicx}  
\usepackage[compress]{cite}
\usepackage{amssymb}
\usepackage{bm} 
\usepackage{dcolumn}
\usepackage{color}
\usepackage{physics}
\usepackage[dvipsnames]{xcolor}
\usepackage{float}
\usepackage{amsfonts}
\usepackage{mathtools}
\usepackage{varioref}
\usepackage{textcomp}
\usepackage{multicol}
\usepackage{enumerate}
\usepackage{tikz}
\usepackage{multicol}
\usepackage{soul}
\usepackage{jcappub}
\usepackage{notoccite}
\usepackage{graphicx}
\usepackage{float}
\usepackage[caption=true]{subfig}
\RequirePackage{parskip}
\usepackage{soul}
\usepackage{pdfrender}

\usepackage{pifont}
\usepackage{ccfonts}
\usepackage{palatino}

\def\RN{Reissner-Nordstr\"{o}m }
\labelformat{section}{Section #1}
\labelformat{subsection}{Section #1}
\labelformat{subsubsection}{Section #1}
\labelformat{subsubsubsection}{Section #1}
\labelformat{equation}{Eq.~(#1)}
\labelformat{figure}{Fig.~[#1]}
\labelformat{subfigure}{Fig.~[\thefigure#1]}
\labelformat{table}{Tab.~#1}
\labelformat{appendix}{Appendix #1}
\definecolor{lime}{HTML}{A6CE39}
\DeclareRobustCommand{\orcidicon}{
	\begin{tikzpicture}
		\draw[lime, fill=lime] (0,0)
		circle [radius=0.16]
		node[white] {{\fontfamily{qag}\selectfont \tiny ID}};
		\draw[white, fill=white] (-0.0625,0.095)
		circle [radius=0.007];
	\end{tikzpicture}
	\hspace{-2mm}
}
\foreach \x in {A, ..., Z}{\expandafter\xdef\csname orcid\x\endcsname{\noexpand\href{https://orcid.org/\csname orcidauthor\x\endcsname}
		{\noexpand\orcidicon}}
}

\newcommand{\B}{ \abs{ \mathbf{B}}}
\newcommand{\BT}{ \abs{ \mathbf{B}_{T}}}
\newcommand{\V}{\mathbf{R}}

\newcommand{\T}{\mathcal{T}}
\newcommand{\p}{\mathfrak{p}}

\title{Exploring Axions through the Photon Ring of a Spherically Symmetric Black Hole}

\author[a]{\bf Sourov Roy,\note{Corresponding author.}}
\author[a]{Pratick Sarkar,}
\author[b,c,1\orcidA{}]{Subhadip Sau}
\author[a]{and Soumitra SenGupta}

\affiliation[a]{School of Physical Sciences, Indian Association for the Cultivation of Science,\\2A \& 2B Raja S.C Mullick Road, Kolkata-700032, India}
\affiliation[b]{Department of Physics, Jhargram Raj College,\\Jhargram, West Bengal-721507, India}
\affiliation[c]{Institute of Astronomy, Space and Earth Science (IASES),\\Bidhan Sishu Sarani, Kolkata-700054, India}

\emailAdd{tpsr@iacs.res.in}
\emailAdd{spsps2523@iacs.res.in}
\emailAdd{subhadipsau2@gmail.com}
\emailAdd{tpssg@iacs.res.in}

\abstract{ 
	In this study, we examine the phenomenon of photon axion conversion occurring in the spacetime surrounding a black hole. Specifically, we focus on the potential existence of a magnetic field around the supermassive black hole M87*, which could facilitate the conversion of photons into axions in close proximity to the photon sphere. While photons traverse through the curved spacetime, they spend time near the photon sphere, where conversion of these photons into axions takes place. Consequently, this process leads to a decrease in the intensity of the black hole's photon ring. To explore the possibilities of detecting these hypothetical axion particles, we propose observing the photon sphere using higher resolution telescopes. By doing so, we can gain valuable insights into the conversion mechanism as well as the nature of the spherically symmetric black hole geometry. Moreover, we also investigate how the photon ring luminosities are affected if the black hole possesses a charge parameter. For instance apart from U(1) electric charge, the presence of extra dimension may induce a {\em tidal charge} with a characteristic signature. It is important to note that the success of the conversion mechanism relies on the axion-photon coupling and mass. As a result, the modified luminosity of the black hole's photon ring offers a valuable means of constraining the axion's mass and coupling parameter within a certain range. Thus our findings contribute to a better understanding of photon axion conversion in the environment of a black hole spacetime and helps us explore the possible existence of extra spatial dimension.
}

\begin{document}
	
	\maketitle
	\flushbottom
	
	\section{Introduction}
	Axions were initially proposed as an elegant resolution to the strong $CP$ problem in quantum chromodynamics (QCD) \cite{Peccei:1977hh, Weinberg:1977ma, Wilczek:1977pj, Shifman:1979if, Zhitnitsky:1980tq, Dine:1981rt,Kim:1979if,Preskill:1982cy,Abbott:1982af,Dine:1982ah}. These hypothetical pseudoscalar particles have also been considered as a potentially promising candidate for dark matter \cite{Duffy:2009ig,Hui:2016ltb, Chadha-Day:2021szb,Adams:2022pbo} due to their characteristics, abundance in universe and incredibly weak coupling. It is also thought that axion-like particles may be essential to our comprehension of inflation since these are identified with the Nambu-Goldstone boson of the spontaneously broken shift symmetry\cite{Freese:1990rb,Adams:1992bn,Kim:2004rp,Kaloper:2008fb,Dimopoulos:2005ac}. The mass of QCD-axions is associated with their coupling constant, while beyond-the-standard-model theories, such as higher-dimensional theories, predict the existence of axion-like particles (ALPs) with uncoupled mass and coupling constant \cite{Svrcek:2006yi, Arvanitaki:2009fg}. Studies of physical, astrophysical and cosmological effects of axions have already been done in literature\cite{Berezhiani:1989fp,Khlopov:1999tm,Sakharov:1996xg}. Evidently, examining coupling strength of axions and their mass is an important area of research.
	
	Axions interact with photons through the coupling term $\mathcal{L}^{int} = -(\frac{g_{\Phi\gamma}}{4}) \Phi F_{\mu \nu}\tilde{F}^{\mu\nu}$, where $g_{\Phi\gamma}$ represents the axion-photon coupling constant, $\Phi$ denotes the axion field, $F_{\mu\nu}$ is the electromagnetic field strength tensor, and $\tilde{F}^{\mu\nu}$ is its dual. This interaction makes it possible for axions to become photons and vice versa when there is an external magnetic field present\cite{Maiani:1986md, PhysRevD.37.1237}. The phenomenon of photon-axion conversion is foundational for the search of solar axions \cite{Armengaud:2014gea, CAST:2017uph} and axion dark matter \cite{ADMX:2009iij}. It has also been proposed as a potential explanation for supernova dimming \cite{Csaki:2001yk,Csaki:2001jk, Deffayet:2001pc, Grossman:2002by}, and could potentially lead to spectral distortions in the cosmic microwave background \cite{Mirizzi:2009nq,Tashiro:2013yea}. The detection of high energy gamma-ray signals\cite{Galanti:2022pbg,Troitsky:2022xso,Baktash:2022gnf,Lin:2022ocj,Gonzalez:2022opy,Nakagawa:2022wwm,Carenza:2022kjt,Wang:2023okw} and $X$-ray or gamma rays from sources like active galactic nuclei \cite{Hooper:2007bq, Hochmuth:2007hk,DeAngelis:2007wiw,HESS:2013udx,Fermi-LAT:2016nkz,Marsh:2017yvc,Zhang:2018wpc,Reynolds:2019uqt} is based on re-conversion of  photons into axions in the extragalactic space such that the photons can escape the electron-positron pair production\cite{Simet:2007sa,Mirizzi:2009aj,Meyer:2013pny}. However, despite various scenarios, the lack of observational evidence imposes constraints on the axion coupling constant \cite{Dolan:2022kul, Dessert:2022yqq}.
	
	Strong field gravitational interaction, such as that occurring near a black hole's horizon, is anticipated to reveal a multitude of details about the basic properties of background geometry. Direct observations exploring the near horizon zone of a black hole geometry were not available until recently. The chances of understanding the nature of strong gravity have significantly improved with the two subsequent ground-breaking discoveries, namely the gravitational wave measurements from the collision of binary black holes and neutron stars\cite{LIGOScientific:2016aoc,LIGOScientific:2017vwq} and imaging the shadow of the supermassive black holes at the centre of the M87 galaxy\cite{Fish:2016jil,EventHorizonTelescope:2019dse,EventHorizonTelescope:2019uob,EventHorizonTelescope:2019jan,EventHorizonTelescope:2019ths,EventHorizonTelescope:2019pgp,EventHorizonTelescope:2019ggy}. Recent observations of the black hole in the center of the M87 galaxy (M87$^*$) through the Event Horizon Telescope imaged polarized synchrotron emission at 230 GHz on event horizon scales\cite{Akiyama_2021}. This polarized synchrotron radiation provides insights into the structure of magnetic fields and the properties of the plasma near the event horizon. Based on these observations, the EHT collaboration estimated the magnetic field strength to be around 1-30 Gauss, the average electron density ($n_{e}$) to be approximately $10^{4-7}$ cm$^{-3}$, and the electron temperature ($T_e$) of the radiating plasma to be about $10^{11}$ K \cite{EventHorizonTelescope:2021srq}. Further, One can find the probing techniques of superradiant axion using photon ring birefringence with EHT data in \cite{Chen:2019fsq,Chen:2022oad,Chen:2022kzv}.
	
	So it is interesting to investigate the photon-axion conversion around black holes. The wavelengths of both photons and axions are such that the magnetic field persists over a larger radial distance, facilitating a smooth conversion process. When photons pass near a black hole, they experience gravitational bending caused by the strong gravity even in presence of extra dimension\cite{Li:2020drn,Hou:2021okc}. Consequently, these photons begin to orbit the black hole due to its powerful gravitational pull. Photons with an impact parameter below a critical value ($b_c$) will enter the black hole's event horizon, while those with exactly the critical impact parameter will move in an unstable circular orbit around the black hole which eventually results in a bright ring. The conversion of photons to axions reduces the number of
	photons escaping the photon sphere, resulting in a dimming effect on the bright ring. Thus, the presence of this conversion leads to  darkening of the photon ring in the observed image. The nature of the gravitational theory can also be better understood by looking at the spectrum of photons, originating from the near horizon region.  Dimming of the photon ring has been extensively  studied for M87* black hole assuming it to be a Schwarzschild black hole\cite{Nomura:2022zyy}.  In this article, we have provided the technique to predict the dimming rate in various alternative theories of gravity for non-rotating black holes. We have used our method to find the signature of extra-dimension by observing the dimmed photon ring.
	
	The paper is structured as follows: In \ref{Sec:PH_AX_CONV},  the  mechanism of photon-axion conversion has been described. Moving on to \ref{Sec:BH_PH}, we explore the time it takes for photons to orbit around the photon sphere of the black hole. The source of the photons near the galactic black holes and  the impact parameter of these photons in terms of their emission angle has been reviewed in \ref{Sec:PH_Near_BH}. The next section i.e. \ref{Sec:PH_DIM} has been dedicated to the discussion on the axion-photon conversion probability in the spherically symmetric spacetime along with the effect of extra dimension on the dimming of the photon ring of the black hole. Concluding remarks with a discussion on major findings have been provided in \ref{Sec:Conclusion}. In this paper, we set $\hbar=c=G=k_{B}=1$, where $\hbar$ is the reduced Planck constant, $c$ is the speed of light, $G$ is the gravitational constant, and  $k_{B}$ is the Boltzmann constant.

	\section{Photon-Axion Conversion Mechanism}\label{Sec:PH_AX_CONV}
	In this section, we study the conversion between photons and axions in a constant external magnetic field in a flat space.
	We consider the following photon-axion system\cite{PhysRevD.37.1237,Masaki:2017aea,Hochmuth:2007hk,PhysRevD.72.023501}
	\begin{align}\label{Eq:action_ph_ax}
		\mathcal{S} = &\, \int \dd^4 x\left(- \frac{1}{4} F_{\mu\nu} F^{\mu\nu} - \frac{1}{2} \partial_\mu \Phi \partial^\mu \Phi - \frac{1}{2} m_\Phi ^2 \Phi^2 \right. \left. - \frac{1}{4} g_{\Phi\gamma} \Phi F_{\mu\nu} \widetilde{F}^{\mu\nu}\right) 
	\end{align}
	where $\Phi$ is the pseudoscalar field (axion) with mass $m_{\Phi}$, $g_{\Phi \gamma}$ is the axion-photon coupling constant, $F_{\mu \nu}\equiv (\partial_{\mu}A_{\nu}-\partial_{\nu}A_{\mu})$ is the electromagnetic field strength tensor for electromagnetic gauge field $A_{\mu}$. The dual of $F_{\mu\nu}$ is  represented by $\widetilde{F}_{\mu\nu}\equiv \frac{1}{2}\epsilon_{\mu \nu \rho \sigma}F^{\rho\sigma} $ with $\epsilon_{\mu \nu \rho \sigma}$ being completely anti-symmetric tensor in its indices. The fourth term in the action represents the $CP$-conserving interaction between the axion and the electromagnetic field. Such interactions may arise in the context of one of the solutions of strong CP problem in QCD as well as the requirement for canceling gauge anomaly in a string inspired model through the resulting electromagnetic Chern-Simons term\cite{PhysRevLett.89.121101,Maity:2003im,Antoniadis:2006wp,Maity:2007un}.
	From the action given in \ref{Eq:action_ph_ax}, we have the Klein Gordon equation for the axion field as 
	\begin{align}\label{phi_equation}
		(\Box-m_\Phi ^2)\Phi=\frac{1}{4}g_{\Phi \gamma}F_{\mu\nu}\widetilde{F}^{\mu \nu}
	\end{align}
	and the Maxwell's equations are given by
	\begin{align}\label{maxwell_eq}
		\partial_\mu F^{\mu\nu}=-g_{\Phi\gamma}\widetilde{F}^{\rho\nu}\partial_\rho \Phi
	\end{align}
	where we have used the Bianchi identity $\partial_{\mu}\widetilde{F}^{\mu\nu}=0$ to derive the equations. Needless to say, although the Bianchi identity remains unchanged, Maxwell's equations get modified due to the presence of the photon-axion coupling term.
	
	We analyse a scenario, in which electromagnetic waves spread against a constant magnetic field $\mathbf{B}$. This is essential because this magnetic field will induce the photon-axion conversion phenomenon, as we will show in this section. The magnetic field and electromagnetic waves combine to form the electromagnetic field
	\begin{align}
		F_{\mu \nu}=\bar{F}_{\mu\nu}+\partial_{\mu}A_{\nu}-\partial_{\nu}A_{\mu}
	\end{align}
	In the simplest scenario,  a constant magnetic field has been considered as our background electromagnetic field such that the background field $\bar{F}_{\mu\nu}$ is represented as
	\begin{subequations}
		\begin{align}
			\bar{F}_{0i}&=\bar{E}_{i}=0\\
			\widetilde{\bar{F}}_{0i}&=\frac{1}{2}\epsilon_{0ijk} \bar{F}_{jk}=B_i
		\end{align}
	\end{subequations}
	We select Coulomb gauge for the propagating photons $A_{\mu}$ as $\bm{\nabla}.\bm{A}=0$ so that the third component of $\bm{A}$ i.e $A_{z}$ vanishes. In the leading approximation with the dispersion relation $ w\simeq k $, we write the 3-vector $\bm{A}$ as a plane wave solution:
	\begin{align}\label{Eq:VP}
		\bm{A}(z,t)\ =\ i \left(
		\begin{array}{c}
			A_{x}(z)\\
			A_{y} (z)\\ 
			0 \\    
		\end{array}
		\right) e^{- i \omega t}\ 
	\end{align}
	assuming that the spatial variation of magnetic fields is substantially larger than that of photon or axion wavelength. If one further assumes the relativistic axion with momentum $k \gg m_{\Phi}$, the solution of $\Phi$ field can be written as
	\begin{flalign}\label{Eq:sF}
		\Phi(z,t)=\Phi(z)e^{-i\omega t}
	\end{flalign}

\begin{figure}[h]
		\centering
		\includegraphics[width=8cm]{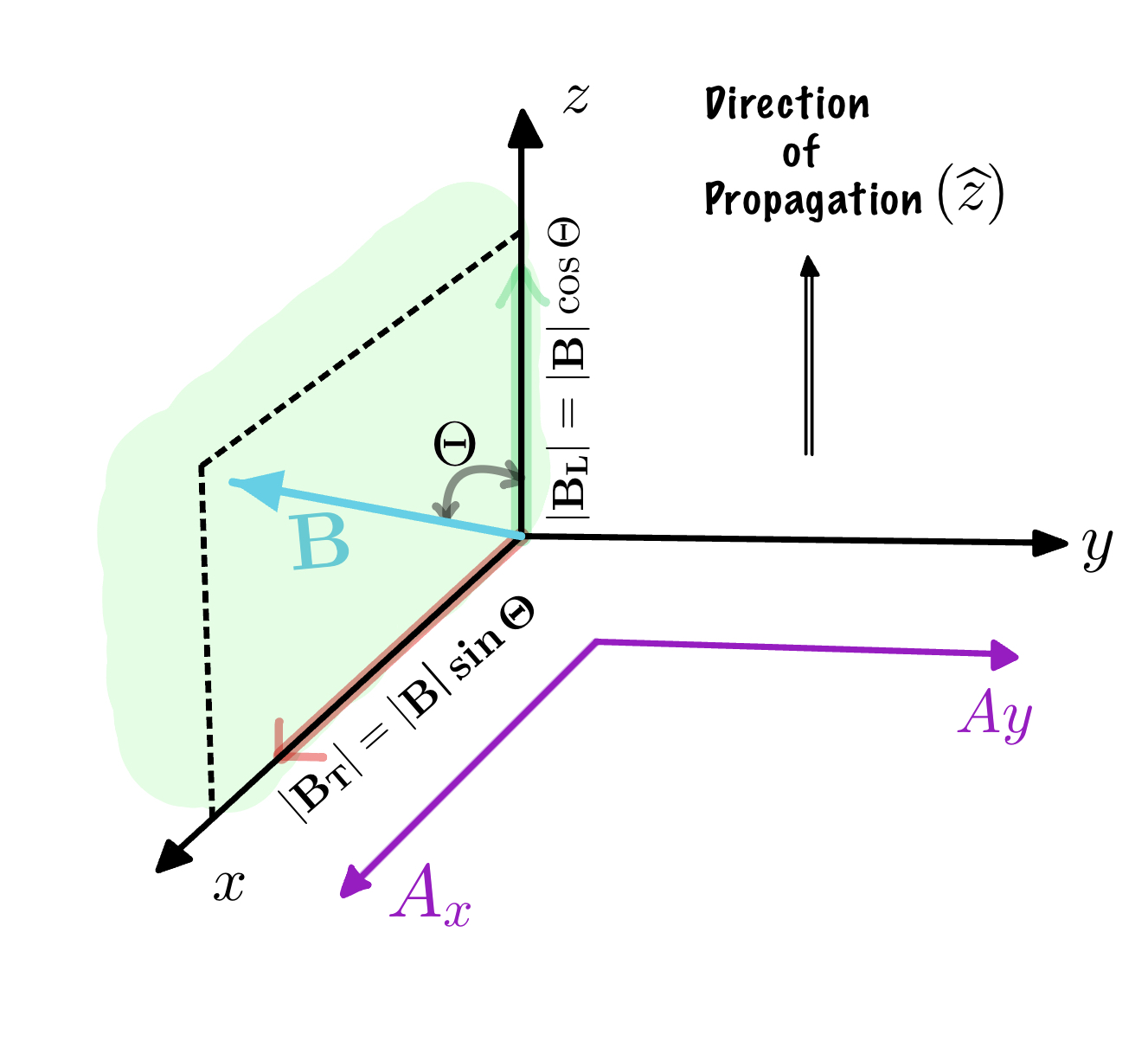}
		\caption{Direction of magnetic field has been shown in the cartesian coordinate system. The magnetic field $\bf{B}$ lies in the $x$-$z$ plane making an angle $\Theta$ with the propagation vector of the photon. The transverse component of the magnetic field is along the $x$-axis and is denoted by $\mathbf{B}_{T}$.  This component of the magnetic field is parallel to the component $A_{x}$, whereas $A_{y}$ is perpendicular to the  transverse part of the magnetic field.} 
		\label{fig:mag_field_config}
	\end{figure}

	We consider a monochromatic light beam traveling along $z$ direction. Magnetic field $\bm{B}$ lies in the $x-z$ plane as shown in \ref{fig:mag_field_config}, where $\Theta$ is the angle between the direction of $\bf{B}$ and $z$ axis. $A_{x}$ and $A_{y}$ denote components of the vector potential parallel and perpendicular to $\mathbf{B}_T$ respectively, where $\mathbf{B}_{T}$ denotes the projection of $\bm{B}$ on the  $x$ axis.

	{
		The spatial components of \ref{maxwell_eq} written under Coulomb gauge as}
	\begin{align}\label{maxwell_with_B}
		\Box\bm{A}-\bm{\nabla}\dot{{A}}^0=g_{\Phi\gamma}\mathbf{B}\dot{\Phi}
	\end{align}
	Here, a dot denotes the time derivative.
	Time component of gauge fields satisfies the following equation
	\begin{align}
		\bm{\nabla}^2 {A}^0=-g_{\Phi\gamma}\mathbf{B}.\bm{\nabla}\Phi
	\end{align}
	The equation of motion of the axion can be rewritten as
	\begin{align}\label{Eq:KG}
		\left(\Box-m_\Phi ^2\right)\Phi=-g_{\Phi\gamma}\mathbf{B}.(\dot{\bm{A}}+\bm{\nabla}A^0)
	\end{align}
	It is clear from \ref{maxwell_with_B} and \ref{Eq:KG} that only the component of $\bm{A}$ parallel to $\mathbf{B}$ mixes with the axion.
	We have assumed $\mathbf{B}$ to be in the $x-z$ plane without losing generality. Now we have the following components for the magnetic field and propagating photons in $(x,y,z)$ coordinate system:
	\begin{flalign}
		\mathbf{B}&=({{\B}}\sin\Theta, 0, {{\B}}\cos\Theta)\\
		\mathbf{A}&=(iA_{x}(z), iA_{y}(z), 0)e^{-i\omega t}\label{Eq:Ph_Field}
	\end{flalign}
	where $\Theta$ is the angle between the direction of $\mathbf{B}$ and $z$ axis. $A_{x}$ and  $A_{y}$ denote components of the photon field along the $x$ axis and $y$ axis respectively. 
	The vacuum polarizability effect due to photon-photon interaction, in the limit where photon frequencies are small in comparison to electron mass and magnetic field strengths are weak in comparison to critical field strengths, in the form of Euler-Heisenberg effective Lagrangian is given by\cite{Mirizzi:2006zy}
	\begin{align}
		\mathcal{L}_{EH}\ =\ \frac{\alpha^2}{90m^4_e}
		\left[\left(F_{\mu\nu}F^{\mu\nu}\right)^2+\frac{7}{4}\left(\widetilde{F}_{\mu\nu}F^{\mu\nu}\right)^2 \right]\ ,
	\end{align}
	where $\alpha=\left(\dfrac{1}{137}\right)$ is the fine structure constant and $m_e\equiv511 ~$ keV is the electron mass. This Lagrangian, arises due to the quantum field theoretic effect describing the one-loop corrections to classical electrodynamics, contributing the following term in the right-hand side of  \ref{maxwell_eq}
	\begin{flalign}
		\dfrac{4\alpha^{2}}{45m_{e}^{4}}\partial_{\mu}\left(F_{\alpha\beta}F^{\alpha\beta}F^{\mu\nu}+\dfrac{7}{4} F_{\alpha\beta}\widetilde{F}^{\alpha\beta}\widetilde{F}^{\mu\nu} \right)
	\end{flalign}
	Using \ref{Eq:Ph_Field} and assuming plane wave solution, we have the following equations in the linear order of $\bm{A}$
	\begin{align}
		F_{\alpha\beta}F^{\alpha\beta}&=2\B^2 +4\omega \B \sin\Theta A_{y}\\
		F_{\alpha\beta}\widetilde{F}^{\alpha\beta}&=-4\omega \B \sin\Theta A_{x}
	\end{align}
	The Euler-Heisenberg Lagrangian modifies Maxwell's equation by the term 
	\begin{equation}
		\frac{4\alpha^2}{45 m_e ^4}\partial_\mu\left[\left(2\B^2+4\omega \B \sin\Theta A_{y}\right)F^{\mu\nu}+\frac{7}{4}\left(-4 \omega \B \sin\Theta  A_{x}\right)\widetilde{F}^{\mu\nu}\right]
	\end{equation}
	With this term added to the right side of \ref{maxwell_with_B} we have
	\begin{align}
		\Box A_{x}+7 \omega^2 \xi \sin^2 \Theta A_{x}+\omega g_{\Phi \gamma}\B \sin\Theta \Phi=0   \\
		\Box A_{y}+4 \omega ^2 \xi \sin^2 \Theta A_{y}=0
	\end{align}
	where,
	\begin{equation}
		\xi=\frac{\alpha}{45\pi}\left(\frac{\B}{\B_{\rm crit}}\right)^2,\quad \B_{\rm crit}\equiv\frac{m_e ^2}{e}
	\end{equation}
	
	By adding a term $-\omega_{\rm pl}^2 A_{x}$ to the equations of motion, the impacts of the surrounding plasma can be taken into account, where $\omega_{\rm pl}$ is plasma frequency,
	\begin{flalign}
		\omega_{\rm pl}\equiv \sqrt{\dfrac{4\pi\alpha n_{e}}{m_{e}}}=3.7\times 10^{-11} {\rm eV}\sqrt{\dfrac{n_{e}}{\rm cm^{-3}}}
	\end{flalign}
	Here $n_{e}$ is the number density of the electron in the medium in which the photon propagates.
	Here $m_{e}$ is the mass of the electron and $\alpha$ is the fine-structure constant.
	The $A_{x}$ component only mixes with the axion, so the relevant equations are
	\begin{subequations}\label{Eq:Field_sol}
		\begin{flalign}
			\Box A_{x}-\omega_{\rm pl}^2 A_{x}+7 \omega^2 \xi \sin^2 \Theta A_{x}+\omega g_{\Phi \gamma}\B \sin\Theta \Phi&=0 \\
			(\Box-m_\Phi ^2)\Phi +\omega g_{\Phi \gamma}\B \sin\Theta A_{x}&=0
		\end{flalign}
	\end{subequations}
	The solutions of the fields  as given in \ref{Eq:VP} and \ref{Eq:sF} are now expressed as
	\begin{subequations}
		\begin{flalign}
			A_{x}(t,z)&=\widetilde{A}(z) e^{-i(\omega t-kz)}+ \rm h.c\\
			\Phi(t,z) &=\widetilde{\Phi}(z) e^{-i(\omega t-kz)}+\rm h.c
		\end{flalign}
	\end{subequations}
	The photons propagating over a distance in the $z$-direction convert into axions and their amplitudes also vary with the distance $z$. It is anticipated that these amplitudes would vary slowly in the sense that the ratio of the second-order partial derivative and the first-order partial derivative of the amplitudes with respect to the propagation direction $z$ is much less than the momentum $k$ associated with the fields i.e $|\partial_z ^2 \widetilde{A}(z)|\ll k |\partial_z \widetilde{A}(z)|$ and $|\partial_z ^2 \widetilde{\Phi}(z)|\ll k |\partial_z \widetilde{\Phi}(z)|$.
	
	Assuming the ultra-relativistic limit $|m_\phi|^2 \ll w^2$ and $\omega \sim k$ the operator,
	\begin{align}
		\Box=(\omega+i\partial_z)(\omega-i\partial_z)=(\omega+i\partial_z)(\omega+k)\approx 2\omega(\omega+i\partial_z)
	\end{align}
	With  this lowest-order approximation, we have
	\begin{subequations}
		\begin{flalign}
			\Box A_{x}(t,z)\simeq 2i \omega \partial_z \widetilde{A}(z) e^{-i(\omega t-kz)}+\rm h.c.\\
			\Box \Phi(t,z)\simeq 2i \omega \partial_z \widetilde{\Phi}(z) e^{-i(\omega t-kz)}+\rm h.c.
		\end{flalign}
	\end{subequations}
	The equations of motion reduce to
	\begin{subequations}
		\begin{flalign}
			i\frac{d}{dz}\widetilde{A}(z)&=\left(\frac{\omega_{\rm pl}^2}{2w}-  \frac{28\alpha^2 \omega}{90 m_e ^4}\left(\B \sin \Theta \right)^2\right)\widetilde{A}(z)-\frac{1}{2}g_{\Phi\gamma}\B\sin\Theta \widetilde{\Phi}(z)\\
			i\frac{d}{dz}\widetilde{\Phi}(z)&=\left(-\frac{1}{2}g_{\Phi\gamma}\B\sin\Theta\right)\widetilde{A}(z)+\frac{m_\Phi ^2}{2\omega}\widetilde{\Phi}(z)
		\end{flalign}
	\end{subequations}
	It is convenient to rewrite the equations as
	\begin{equation}\label{diffeq}
		i\frac{d}{dz}\Psi(z)=\mathbf{M}\Psi(z)
	\end{equation}
	where we have used the notations
	\begin{equation}
		\Psi(z)=
		\begin{bmatrix}
			\widetilde{A}(z)\\
			\widetilde{\phi}(z)\\     
		\end{bmatrix}
	\end{equation}
	\begin{equation}
		\bf{M}=
		\begin{bmatrix}
			\Delta_{\parallel} & -\Delta_{\rm M} \\
			-\Delta_{\rm M} & \Delta_\Phi \\
		\end{bmatrix}
	\end{equation}
	with 
	\begin{subequations}
		\begin{flalign}
			\Delta_{\parallel}&=\Delta_{\rm pl}-\Delta_{\rm vac}\\
			\Delta_{\rm pl}&\equiv \dfrac{\omega_{\rm pl}^{2}}{2\omega}=6.9\times 10^{-25} {\rm eV} \left(\dfrac{n_{e}}{\rm cm^{-3}}\right) \left(\dfrac{\rm keV}{\omega} \right) \\
			\Delta_{\rm vac}&\equiv \frac{28\alpha^2 \omega}{90 m_e ^4}\left(\B \sin \Theta \right)^2= 9.3\times 10^{-29} {\rm eV}\left(\dfrac{\omega}{\rm keV}\right)\left(\dfrac{\B}{\rm Gauss}\right)^{2}\sin^{2}\Theta \\
			\Delta_{\rm M}&\equiv \frac{1}{2}g_{\Phi\gamma}\B\sin\Theta = 9.8\times 10^{-23} {\rm eV} \left(\dfrac{g_{\Phi\gamma }}{10^{-11}\rm GeV^{-1}} \right)\left(\dfrac{\B}{\rm Gauss} \right)\sin\Theta \\
			\Delta_\Phi&\equiv\frac{m_\Phi ^2}{2\omega} =5\times 10^{-22} {\rm eV} \left(\dfrac{m_{\Phi}}{\rm neV} \right)^{2}\left(\dfrac{\rm keV}{\omega} \right) 
		\end{flalign}
	\end{subequations}
	These $\Delta_{i}$s (i: pl, vac, M, $\Phi$) have been estimated for our study of M87* and are all defined positive in contrast with the definition  in \cite{PhysRevD.37.1237} . 
	The matrix $\bf{M}$ has eigenvalues that are
	\begin{equation}
		\lambda_{\pm}=\frac{(\Delta_{\parallel}+\Delta_{\Phi})\pm \sqrt{(\Delta_{\parallel}-\Delta_{\Phi})^2 +(2\Delta_{\rm M})^2}}{2}
	\end{equation}
	Now we introduce an orthogonal matrix $\bf{O}$ to diagonalize $\bf{M}$ such that
	\begin{align}
		\mathbf{O}^T \bf{M}\bf{O}=
		\begin{bmatrix}
			\lambda_+ & 0 \\
			0 & \lambda_- \\
		\end{bmatrix}
		,~~
		\bm{O}\equiv  
		\begin{bmatrix}
			\cos \vartheta & \sin \vartheta \\
			-\sin \vartheta & \cos \vartheta \\
		\end{bmatrix}
	\end{align}
	where $\vartheta$ is the \emph{mixing angle} related as
	\begin{align}
		\vartheta =\dfrac{1}{2} \arctan\left( \frac{2 \Delta_{\rm M}}{\Delta_{\Phi} - \Delta_\parallel}\right).
		\label{angle}
	\end{align}\\
	Using the matrix $\mathbf{O}$, \ref{diffeq} reduces to 
	\begin{align}
		i \frac{d}{dz} (\mathbf{O}^T \Psi(z)) = \begin{bmatrix}
			\lambda_+ & 0\\
			0& \lambda_-
		\end{bmatrix}
		(\mathbf{O}^T \Psi(z)).
	\end{align}
	Solving the equation, we have
	\begin{align}
		\Psi(z)
		= \mathbf{O} \begin{bmatrix}
			e^{-i\lambda_+ z} &0\\
			0& e^{-i\lambda_- z} 
		\end{bmatrix}
		\mathbf{O}^T \Psi(0).
	\end{align}
	In the end,  we have found the general solutions as
	\begin{subequations}
		\begin{align}
			\widetilde{A}(z)
			&= \left(\cos^2 \vartheta e^{-i\lambda_+ z} + \sin^2 \vartheta e^{-i\lambda_- z} \right) \widetilde{A}(0)
			+ \sin \vartheta \cos \vartheta \left(e^{-i\lambda_+ z } - e^{-i\lambda_- z} \right) \widetilde{\Phi}(0)
			\\
			\widetilde{\Phi}(z)
			&= \sin \vartheta \cos \vartheta \left(e^{-i\lambda_+ z} - e^{-i\lambda_- z}\right) \widetilde{A}(0) 
			+ \left(\sin^2 \vartheta e^{-i\lambda_+ z} + \cos^2\vartheta e^{-i\lambda_- z}\right) \widetilde{\Phi}(0)
		\end{align}
	\end{subequations}
	Assuming $\widetilde{\Phi}(0) = 0$ and $\widetilde{A}(0) = 1$ as initial axion density is negligibly small with respect to that of photons. With this initial condition, one can obtain the  probability of conversion of  photon into axion as a function of distance $z$ as
	\begin{align}\label{Eq:Probability}
		P_{\gamma \to \Phi}(z)
		&= \left|\widetilde{\Phi}(z)\right|^2 
		\notag \\
		&= \sin^2 2\vartheta\ \sin^2 \left( \frac{\Delta_{\rm osc}}{2}\ z \right)
		\notag \\
		&= \left( \frac{\Delta_{\rm M}}{\Delta_{\rm osc} / 2} \right)^2 \sin^2 \left( \frac{\Delta_{\rm osc}}{2}\ z \right)
	\end{align}\label{conv_formula}
	where we have defined the \emph{oscillation length} $\Delta_{\rm osc}^{-1}$ as 
	\begin{equation}\label{Eq:Osc_length}
		\Delta_{\rm osc}\ \equiv\ \lambda_+-\lambda_-\ =\ \sqrt{(\Delta_{\Phi}-\Delta_\parallel)^2+(2\Delta_{\rm M})^2}\ .
	\end{equation}

	The key parameters for the photon-axion conversion are:
	\begin{enumerate}[(i)]
		\setlength{\itemsep}{0pt}
		\item  frequency of the propagating photons, $\omega$
		\item  magnetic field  which is perpendicular to the  direction  of the photon propagation, $\mathbf{B}_{T}$ 
		\item axion mass, $m_{\Phi}$
		\item  axion-photon coupling, $g_{\Phi \gamma}$
		\item  number density of the electron in the medium, $n_e$.
	\end{enumerate}
	
	The conversion of photons to axions depends critically on determining  $\Delta_M$ using $g_{\Phi\gamma}$ and $\BT$. The conversion process is suppressed by the contributions of finite axion mass ($\Delta_{\Phi}$), plasma contribution ($\Delta_{\rm pl}$), and one-loop corrections of electrons ($\Delta_{\rm vac}$).
	For relativistic axions, $\omega$ is significantly larger than the axion mass ($m_{\Phi}$), and for photons travelling through a medium, $\omega$ is much larger than the plasma frequency ($\omega_{\rm pl}$). At least, three of the following requirements must be met to the current framework to be valid: 
	\begin{enumerate}[(i)]
		\setlength{\itemsep}{0pt}
		\item $\dfrac{\alpha}{45\pi}\left( \dfrac{\B}{\B_{\rm crit}}\right)^{2}\ll 1$ where critical magnetic field is given by $\B_{\rm crit}\equiv\dfrac{m_{e}^{2}}{\sqrt{4\pi\alpha}}=4\times 10^{13} \rm\ Gauss$,
		\item $\omega\gg m_{\Phi}$ for validating the assumption of relativistic axions,
		\item $\omega \gg \omega_{\rm pl}$ in order for photons to move in the surrounding plasma.
	\end{enumerate}
	Further, from \ref{Eq:Probability} and \ref{Eq:Osc_length}, it can be concluded that the conversion probability becomes more effective when
	\begin{flalign}\label{Eq:Eff_Conv}
		\left(\Delta_{\rm pl}-\Delta_{\rm vac}-\Delta_{\Phi}\right)^{2} \ll 4\Delta_{M}^{2}
	\end{flalign}
	For such effective conversion, the characteristic length scale of the conversion can be expressed as
	\begin{align}
		\Delta_{\rm osc}^{-1} &\simeq (2\Delta_M)^{-1}\label{Eq:Eff_Cov} \\
		&=3.4\times 10^{2}\times(2\times10^{9}M_{\odot})\times\left(\frac{\text{Gauss}}{\BT}\right)\left(\frac{10^{-11}\text{ GeV}^{-1}}{g_{\Phi\gamma}}\right) 
	\end{align}
	It can be observed that for $\B\sim10^{1-2}$ gauss, once the conversion length is compared to the Schwarzschild radius of a supermassive black hole with mass $\sim 10^{9–10} M_{\odot}$,  the two become comparable if $g_{\Phi\gamma}\sim 10^{-11}\rm\ Gev^{-1}$. It is worth mentioning here that the supermassive black hole M87* with mass $6.2^{+1.1}_{-0.5}\times 10^{9} M_{\odot}$ \cite{gebhardt2011black} possesses the magnetic field of strength $\sim (1-30)$ gauss \cite{EventHorizonTelescope:2021srq}  in its vicinity. Since photons can remain in the photon sphere of black holes for a while, we can anticipate that the conversion to axions takes place efficiently.
	From \ref{Eq:Eff_Conv}, one can notice that the efficient conversion can occur if these trivial conditions holds:  $\Delta_{\rm pl}\ll \Delta_{M},
	\Delta_{\rm vac}\ll \Delta_{M},$ and $
	\Delta_{\Phi} \ll \Delta_{M}$. These conditions can be realised better in terms of the photon-axion coupling, magnetic field associated with the black hole and mass of axion as
	\begin{subequations}
		\begin{flalign}
			7\times 10^{-3}\left(\dfrac{n_{e}}{\rm cm^{-3}} \right) &\ll \left(\dfrac{g_{\Phi\gamma}}{10^{-11}{\rm GeV^{-1}}} \right)\left(\dfrac{\omega}{\rm keV} \right)\label{Eq:Eff_con_1} \\
			\left(\dfrac{\omega}{\rm keV} \right)\left(\dfrac{\B}{\rm Gauss} \right) &\ll 1.1\times 10^{6}\left(\dfrac{g_{\Phi\gamma}}{10^{-11}\rm GeV^{-1}} \right)\label{Eq:Eff_con_2}\\
			5.1 \left( \dfrac{m_{\Phi}}{\rm neV}\right) & \ll \left(\dfrac{g_{\Phi\gamma}}{10^{-11}{\rm GeV^{-1}}} \right)  \left( \dfrac{\omega}{\rm keV}\right) \left(\dfrac{\B}{\rm Gauss} \right)\label{Eq:Eff_con_3}
		\end{flalign}
	\end{subequations}
	\ref{Eq:Eff_con_1} and \ref{Eq:Eff_con_3} set    the lower bound on the frequency $\omega$ for efficient conversion. On the other hand, \ref{Eq:Eff_con_2} can be used to determine the upper bound of the frequency.
	Even when $\Delta_{\rm pl}$ is comparable to $\Delta_{M}$, under certain scenarios we can have the efficient conversion. The conditions are
	\begin{enumerate}[(i)]
		\item The first case is $\Delta_{\rm pl} \simeq \Delta_{\Phi}$ and $\Delta_{\rm vac}\ll \Delta_{\Phi}$, which in turn can be translated as
		\begin{flalign}
			\left(\dfrac{m_{\Phi}}{\rm neV} \right)^{2} \simeq 1.4 \times 10^{-3} \left(\dfrac{n_{e}}{\rm cm^{-3}}\right)\\
			\left( \dfrac{\omega}{\rm keV}\right)^{2} \left(\dfrac{\B}{\rm Gauss} \right)^{2} \ll 5.4 \times 10^{6} \left( \dfrac{m_{\Phi}}{\rm neV}\right)^{2}
		\end{flalign}
		\item The second case is $\Delta_{pl}\simeq \Delta_{\rm vac} $ and $\Delta_{\Phi}\ll \Delta_{\rm vac}$, which can be re-written as 
		\begin{flalign}
			\left( \dfrac{\omega}{\rm keV}\right)^{2} \left(\dfrac{\B}{\rm Gauss} \right)^{2}  \simeq 7.4 \times 10^{3} \left(\dfrac{n_{e}}{\rm cm^{-3}} \right)\\
			5.4\times 10^{6}\left(\dfrac{m_{\Phi}}{\rm neV} \right)^{2} \ll \left( \dfrac{\omega}{\rm keV}\right)^{2} \left(\dfrac{\B}{\rm Gauss} \right)^{2} 
		\end{flalign}
	\end{enumerate}
	\begin{figure}[h]%
		\centering
		\subfloat[Variation of the conversion probability (without the sinusoidal part) in $\omega-n_{e}$ plane has been shown assuming the axion mass to be $m_{\Phi}=100\ {\rm neV}$, the photon-axion coupling to be $g_{\Phi\gamma}=10^{-11}\ {\rm GeV}^{-1}$ and magnetic field to be $\B =30\ {\rm gauss}$. ]{{\includegraphics[scale=0.65]{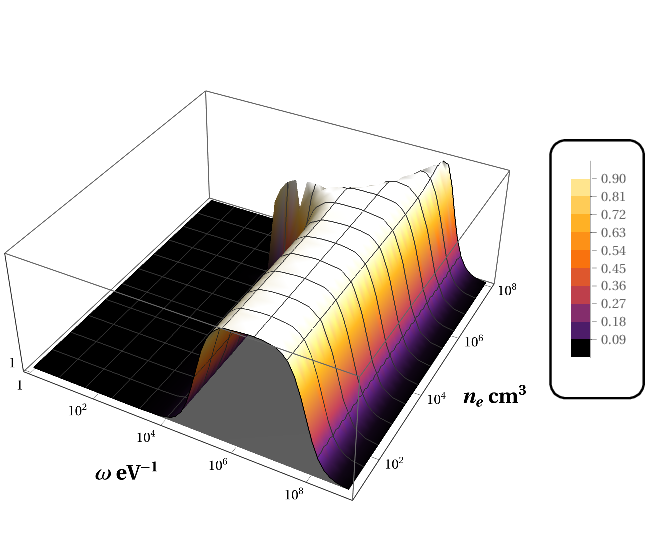} } \label{Fig:DF_1}}
		\qquad
		\subfloat[Variation of the conversion probability (without the sinusoidal part) in $\omega-n_{e}$ plane has been shown assuming the axion mass to be $m_{\Phi}=10\ {\rm neV}$, the photon-axion coupling to be $g_{\Phi\gamma}=10^{-11}\ {\rm GeV}^{-1}$ and magnetic field to be $\B =30\ {\rm gauss}$. ]{{\includegraphics[scale=0.65]{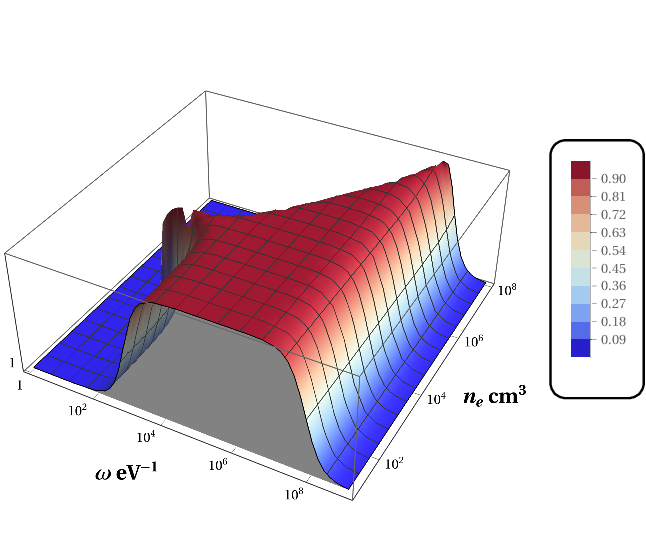} } \label{Fig:DF_2}}
		\qquad
		\subfloat[Variation of the conversion probability (without the sinusoidal part) in $\omega-n_{e}$ plane has been shown assuming the axion mass to be $m_{\Phi}=1\ {\rm neV}$, the photon-axion coupling to be $g_{\Phi\gamma}=10^{-11}\ {\rm GeV}^{-1}$ and magnetic field to be $\B =30\ {\rm gauss}$.] {{\includegraphics[scale=0.65]{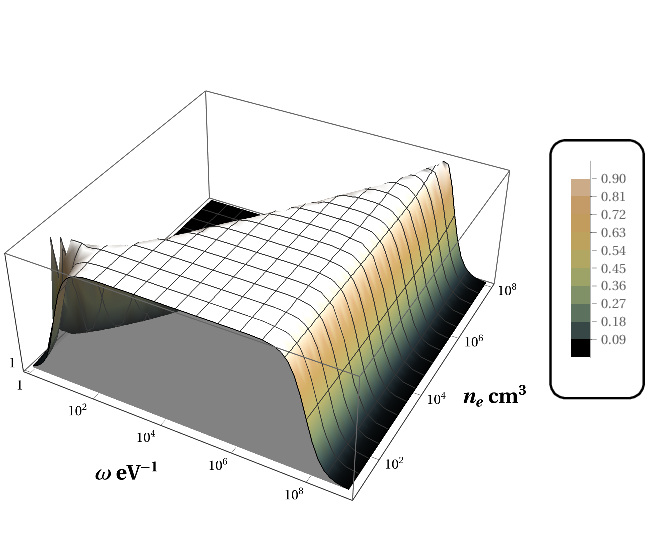} } \label{Fig:DF_3}}
		\qquad
		\subfloat[Density variation of the conversion probability (without the sinusoidal part) has been presented in $\omega-n_{e}$ plane with variation of axion mass in the range $(1-100)\  {\rm neV}$ assuming the photon-axion coupling to be $g_{\Phi\gamma}=10^{-11}\ {\rm GeV}^{-1}$ and magnetic field to be $\B =30\ {\rm gauss}$.]{{\includegraphics[scale=0.65]{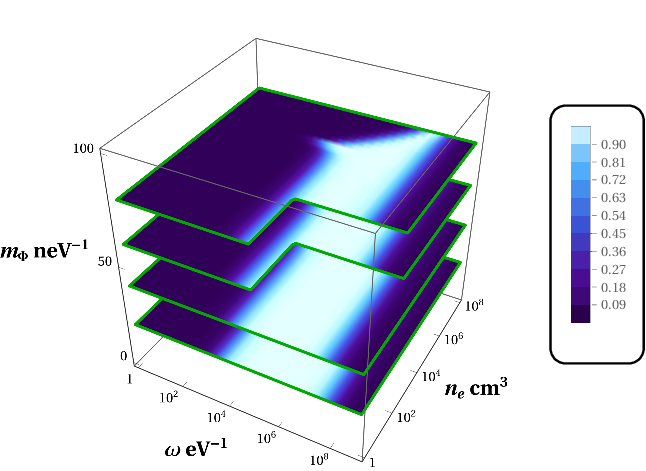}} \label{Fig:DF_4}}
		\caption{ Dependence of the conversion probability (without the sinusoidal part) on the various parameters has been depicted here. The range of the frequency of the photon and the density of the electron have been kept in the range $(1-10^{9})\ {\rm eV}$ and $(1-10^{8})\ {\rm cm^{-3}}$ respectively. Here we have also assumed the magnetic field to be $\B=30\ \rm gauss$ and the axion-photon coupling to be $g_{\phi\gamma}=10^{-11}\ {\rm Gev}^{-1}$.}\label{Fig:DF}%
	\end{figure}
	We aim to visualize the conversion probability in the parameter space, considering the observations from the Event Horizon Telescope. Based on the observations, it is known that the electron number density ranges from approximately $10^4$ to $10^7\  {\rm cm}^{-3}$ and the magnetic field near M87$^*$ is estimated to be around 1-30 gauss\cite{EventHorizonTelescope:2021srq}. We fix the axion-photon coupling to be $10^{-11}$ GeV$^{-1}$ .
	The conversion probability involves two factors: a space-independent factor, $[\Delta_M/(\Delta_{\text{osc}}/2)]^2$, which is unaffected by the spacetime geometry, and a space-dependent oscillation factor that depends on the geometry of the spacetime. 
	
	Only with the space independent factor $[\Delta_M/(\Delta_{\text{osc}}/2)]^2$ in  $\omega$-$n_e$ plane and with axion masses  $m_\Phi=1\ {\rm neV}$, $10\ {\rm neV}$ and $100\ {\rm neV}$, there exists a region where the conversion probability can reach its maximum value as shown in \ref{Fig:DF}. In the case of $m_\Phi=1\ {\rm neV}$ , the plasma effect $\Delta_{\rm pl}$ dominates over the axion mass effect $\Delta_\Phi$ throughout the entire space when $n_e \gtrsim 10^3\ {\rm cm}^{-3}$. This has been shown in \ref{Fig:DF_3}.
	From \ref{Fig:DF_1}, \ref{Fig:DF_2} and \ref{Fig:DF_3}, one can infer that the frequency zone narrows at higher axion masses, such as $m_{\Phi}=10\ {\rm neV}$ or $100\ {\rm eV}$, setting an upper bound on $\omega$ for effective conversion.
	
	In \ref{Fig:DF_4}, the contour plots are stacked vertically for fixed $m_{\Phi}$ values. The horizontal white region extending towards the low-$\omega$ region(upper most one) indicate the resonances where $\Delta_{\rm pl}$ is comparable to $\Delta_{\Phi}$. Additionally, for a fixed value of axion mass, we observe that the white bands become narrower as they extend towards the upper-right portion, connecting to the line of resonance $\Delta_{\rm pl}\sim \Delta_{\rm vac}$.
	Considering that the probability of conversion from photon to axion is influenced by the magnetic field and the electron number density ($n_e$), we will adopt reference values of $\B=30$ gauss for the magnetic field, $n_e\approx 10^{4}$ cm$^{-3}$ for the number density, and an axion mass on the order of $10^{-9}$ eV. At frequencies greater than or equal to $10^{6}$ eV, the creation of electron-positron ($e^{-}-e^{+}$) pairs becomes relevant. However, the photon-axion conversion takes place at lower frequencies, adding a layer of complexity to the discussion.
	\section{Photon orbiting-time around the photon sphere}\label{Sec:BH_PH}
	We take into account the geodesic motion of a photon in a static, spherically symmetric spacetime. The magnetic field strength in this context is on the order of Gauss, and it is not substantial enough to exert any discernible back reaction on the spherically symmetric geometry. The general static, spherically symmetric metric that satisfies Einstien's gravitational field equations is given by
	\begin{equation}\label{Eq:metric_SS}
		ds^2 = -f(r)dt^2 + \frac{1}{g(r)}dr^2 + r^2(d\theta^2 + \sin^2{\theta}d\phi^2)
	\end{equation}
	The geodesic of a photon is represented by $x^{\mu}(\lambda)=(t(\lambda),r(\lambda),\theta(\lambda),\phi(\lambda))$ in this coordinates, where $\lambda$ is affine parameter.  Due to the symmetry of the spacetime, without loss of generality, we consider the geodesics in the equatorial plane, i.e $\theta=\pi/2$.
	The metric given in \ref{Eq:metric_SS} is invariant under time translation and possesses a time-like Killing vector given by $t^{\mu}=(1,0,0,0)$. Similarly, due to the invariance of metric on azimuthal angle gives rise to the space-like Killing vector $\phi^{\mu}=(0,0,0,1)$. The presence of these Killing vectors inspires one to introduce two conserved quantities along the geodesics of both massless and massive particles. These conserved quantities are given by
	\begin{align}
		E&=-(\partial_t)_\mu\frac{dx^\mu}{d\lambda}=f(r)\frac{dt}{d\lambda}\\
		L&=(\partial_\phi)_\mu\frac{dx^\mu}{d\lambda}=r^2\frac{d\phi}{d\lambda}    
	\end{align}
	such that $E$ and $L$ represent the  energy and  angular momentum for massless particles and energy per unit mass and angular momentum per unit mass for massive particles, respectively. In our analysis, we will focus on photon particles i.e. particles with zero rest mass. Hence, from now on, we will use $E$ and $L$ as energy and angular momentum of the photon particle, respectively. The impact parameter $b$ for a photon that approaches a black hole can be expressed in terms of these conserved quantities as	\begin{align}
		b=\dfrac{L}{E}
	\end{align}
	Now, let us consider a dimensionful parameter $s$ such that the four-momentum is given by $p^{\mu}=E \dfrac{dx^{\mu}}{ds}$, where $E$ is the conserved energy. With this definition, the null geodesic satisfies
	\begin{subequations}\label{Eq:mot}
		\begin{align}\label{Eq:Rad_mot}
			\frac{dr}{ds}&=\pm \sqrt{\V(r)} , \hspace{0.5cm}\V(r)=h(r)\left( 1-V_{\rm eff}\right), \hspace{0.5cm}\\
			\dfrac{d\phi}{ds}&=\dfrac{b}{r^{2}}\label{Eq:Rot_mot}\\
			\dfrac{dt}{ds}&=\dfrac{1}{f(r)}
		\end{align}
	\end{subequations}
	where we have defined
	\begin{subequations}
		\begin{flalign}
			h(r)&=\dfrac{g(r)}{f(r)}\\
			V_{\rm eff}&=\dfrac{b^2}{r^2}f(r)
		\end{flalign}
	\end{subequations}
	One can note that for Schwarzschild and \RN spacetime, we have $h(r)=1$.
	
	The photon sphere is the radius at which photons have zero radial velocity, allowing them to revolve around a central gravitating object with a fixed radial distance. This radius of photon sphere can be determined with the following conditions: $\V(r)=0=\dfrac{d \V(r)}{d r}$.
	Let us assume that	the photon orbits have a photon sphere at $r=r_{ph}$, which can be determined by the following algebraic  equation
	\begin{flalign}
		r_{ph}f'(r_{ph})-2 f(r_{ph})=0
	\end{flalign}
	The critical impact parameter for the photons of zero radial  velocity at the photon sphere is given by
	\begin{flalign}
		b_{c}=\dfrac{r_{ph}}{\sqrt{f(r_{ph})}}
	\end{flalign}
	Now, we will study the behaviour of photon geodesics near the photon sphere in perturbative approach\cite{Gralla:2019xty}.
	We define dimensionless fractional deviations around the radius of the photon sphere and the critical impact parameter as follows
	\begin{subequations}
		\begin{align}
			r&=r_{ph}(1+\delta r)\\
			\hspace{0.5cm}b&=b_c(1+\delta b)
		\end{align}
	\end{subequations}
	Expanding the impact parameter around its critical value we have
	\begin{align}
		b 
		=b_c\bigg[1&+\left(\frac{1}{2}-\frac{1}{4}b_c^2 f''(r_{ph})\right)\delta r^2+\left(-\dfrac{1}{12}b_{c}^{2}r_{ph} f'''(r_{ph})+\dfrac{1}{2}b_{c}^{2}f''(r_{ph})-1\right)\delta r^{3}+\mathcal{O}(\delta r^4)\bigg]
	\end{align}
	For our purposes, i.e., orbits close to the photon sphere, $\delta r$ and $\delta b$ must be brought simultaneously to zero at a rate $\delta b \propto \delta r^{2}$. This approximation, which keeps the leading term, offers the form of $\V(r)$ around the photon sphere as 
	\begin{align}
		\V(r) \approx h(r_{ph})\left\{\left[2r_{ph}\frac{f'(r_{ph})}{f(r_{ph})}-\frac{1}{2}r_{ph}^2\frac{f''(r_{ph})}{f(r_{ph})}-3\right]\delta r^{2} -2\delta b\right\}\label{Eq:Pot_approx}
	\end{align}
	From now we define
	\begin{equation}
		2r_{ph}\frac{f'(r_{ph})}{f(r_{ph})}-\frac{1}{2}r_{ph}^2\frac{f''(r_{ph})}{f(r_{ph})}-3=1-\dfrac{1}{2}r_{ph}^{2}\dfrac{f''(r_{ph})}{f(r_{ph})} \equiv \p
	\end{equation}
	and
	\begin{align}
		\left[\frac{1}{2}-\frac{1}{4}b_{c}^2 f^{''}(r_{ph})\right]\left(\frac{M}{r_{ph}}\right)^2 \equiv \mathfrak{a}
	\end{align}
	The turning point is determined by setting the potential to zero. With the leading order approximation near the photon sphere, we have  
	\begin{equation}
		\delta r_{\rm turn}=\sqrt{\frac{2\delta b}{\p}}
	\end{equation}
	Now, we inquire how much affine parameter gathers in an area close to the photon orbit specified by
	\begin{equation}
		-\delta R<\delta r < \delta R
	\end{equation}
	with the assumption  $0<\delta R \ll 1$.\\
	Let us first consider the case when $\delta b <0$. \ref{Eq:Pot_approx} suggests that for $\delta b<0$ there exists no turning point. In this particular case, from \ref{Eq:Rad_mot} we  get
	\begin{align}
		\Delta s& =\int_{-\delta R}^{\delta R}\frac{r_{ph}~ \dd\delta r}{\sqrt{h(r_{ph})}\sqrt{\p \delta r^2-2\delta b}}\\
		& =\frac{r_{ph}}{\sqrt{\p}\sqrt{ h(r_{ph})}} \ln\left[\frac{\p\delta R^2}{-2\delta b}\left(1+\sqrt{1-\frac{2 \delta b}{\p \delta R^2}} ~\right)^2\right]\label{Eq:Log_Div}
	\end{align}
	From \ref{Eq:Log_Div} it is evidently distinct that there exists a logarithmic divergence as $\delta b$ approaches zero. 
	
	Now let us consider the case $\delta b>0 $. In this scenario we have a single turning point. In such a scenario, we have to take $\delta R>\delta r_{turn}$ for the photon to enter the region of interest.	When $\delta b>0$, the trajectories can reach infinity and involve outer turning points and for that we have
	\begin{flalign}
		\triangle s &=2 \times \dfrac{r_{ph}}{\sqrt{h(r_{ph})}}\int_{\sqrt{\frac{2}{\p}\delta b}}^{\delta R}\frac{\dd\delta r}{\sqrt{\p \delta r^2-2\delta b}}\\
		& =\frac{r_{ph}}{\sqrt{\p}\sqrt{h(r_{ph})}} \ln\left[\frac{\p\delta R^2}{2\delta b}\left(1+\sqrt{1-\frac{2 \delta b}{\p \delta R^2}} \right)^2\right]\\
		&=\frac{r_{ph}}{\sqrt{\p}\sqrt{h(r_{ph})}}\ln \left[\frac{\p\delta R^2}{2|\delta b|}\left(1+\sqrt{1-\frac{2 \delta b}{\p \delta R^2}} \right)^2\right] \Theta\left(\delta R^2-(2/\p)\delta b\right)
	\end{flalign}
	When $\delta b\ll (\p/2)\delta R^2$,we have\\
	\begin{equation}\label{Eq:lapse}
		\triangle s\approx \frac{r_{ph}}{\sqrt{\p}\sqrt{h{(r_{ph})}}}\ln\left[2\p\frac{\delta R^2}{\delta b}\right]
	\end{equation}
	Using \ref{Eq:lapse} one can find the lapse in $t$ near the photon sphere, which is given by
	\begin{align}
		\triangle t &\simeq\frac{1}{f(r_{ph})}\frac{r_{ph}}{\sqrt{\p}\sqrt{h(r_{ph})}}\ln\left[2\p\frac{\delta R^2}{\delta b}\right]\\
		&=-\frac{1}{f(r_{ph})}\frac{r_{ph}}{\sqrt{\p}\sqrt{h(r_{ph})}}\ln\left[\frac{b-b_c}{2 \p b_c \delta R^2}\right]\\
		& =-\frac{1}{f(r_{ph})}\frac{r_{ph}}{\sqrt{\p}\sqrt{h(r_{ph})}}\ln\left[\frac{2(b-b_c)}{\p b_c}\times \frac{r_{ph}^2}{\epsilon^2 M^2}\right] \equiv \T(b)
	\end{align}
	where we have redefined the small parameter  as 
	\begin{flalign}
		\epsilon = \left(\dfrac{r_{ph}}{M}\right)\delta r
	\end{flalign}
	Here $M$ is the mass parameter of the given spacetime. 
	
	\section{Photons approaching the photon sphere}
	\label{Sec:PH_Near_BH}

	\subsection{Relation between emission angle and impact parameter}
	
	\begin{figure}[h]
		\centering
		\includegraphics[scale=0.7]{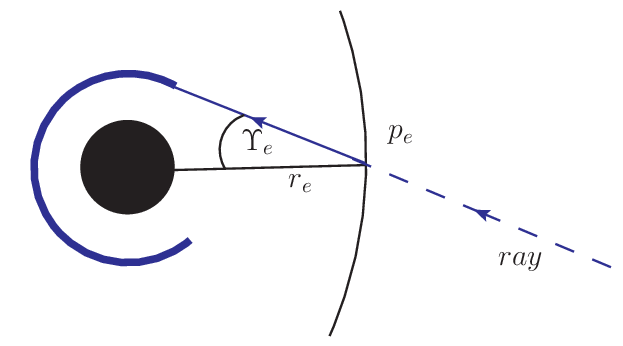}
		\caption{ The blue curve represents a light ray emitting from a point $p_e$ toward the photon sphere. The point $p_e$ is positioned on a sphere with a radius labeled as $r_e$  and it is centred around the black hole within a spherically symmetric coordinate system. The angle $\Upsilon_e$ is the zenith angle.}
		\label{Fig:LT_emit}
	\end{figure}
	
	Consider a light beam that is directed towards a black hole photon sphere with an impact parameter of $b$ from a point $p_e$ that is situated at $r_e$ in spherically symmetric spacetime coordinates. As illustrated in \ref{Fig:LT_emit},  we introduce an angle $\Upsilon_e$ between the initial direction of the incident photon and the direction to the black hole's centre, in order to characterise the track of the photon. For spherically symmetric spacetime as given in \ref{Eq:metric_SS} and assuming $g(r)=f(r)$  the tetrads are given by
	\begin{subequations}	
		\begin{flalign}
			e^{(0)}&=\left(\sqrt{f(r)},0,0,0\right)\\
			e^{(1)}&=\left(0,\frac{1}{\sqrt{f(r)}},0,0\right)\\
			e^{(2)}&=\left(0,0,r,0\right)\\
			e^{(3)}&=\left(0,0,0,r\sin\theta \right)
		\end{flalign}
	\end{subequations}
	The tetrads $e^{(1)}$ and $e^{(3)}$ are the representative of  orthonormal bases parallel and normal to the direction to the center of the black hole. Hence, the angle $\Upsilon_e$ is given by
	\begin{equation}
		\tan \Upsilon_e =\abs{\frac{k^\mu e^{(3)}_\mu}{k^\mu e^{(1)}_\mu}}_{p_e}=\abs{r\sqrt{f(r) }\sin\theta\frac{d\phi}{dr}}_{p_e}
	\end{equation}
	where $k^\mu=\dfrac{dx^\mu}{d\lambda}$ is the tangent vector to the geodesic of the photon with the  affine parameter $\lambda$. For simplicity, we assume that  the plane of the geodesic is at $\theta=\pi/2$ plane. Using \ref{Eq:mot}, one can find 
	\begin{equation}
		b=\frac{r_{e}}{\sqrt{f(r_e)}}\sin\Upsilon_e
		\label{B6}
	\end{equation}
	\ref{B6} enables us to relate the emission angle $\Upsilon_{e}$, defined at a point $p_{e}$, with the impact parameter.	
	\subsection{Flow into photon sphere from spherical region}
	
\begin{figure}[h]
		\centering
		\includegraphics[width=0.7\textwidth]{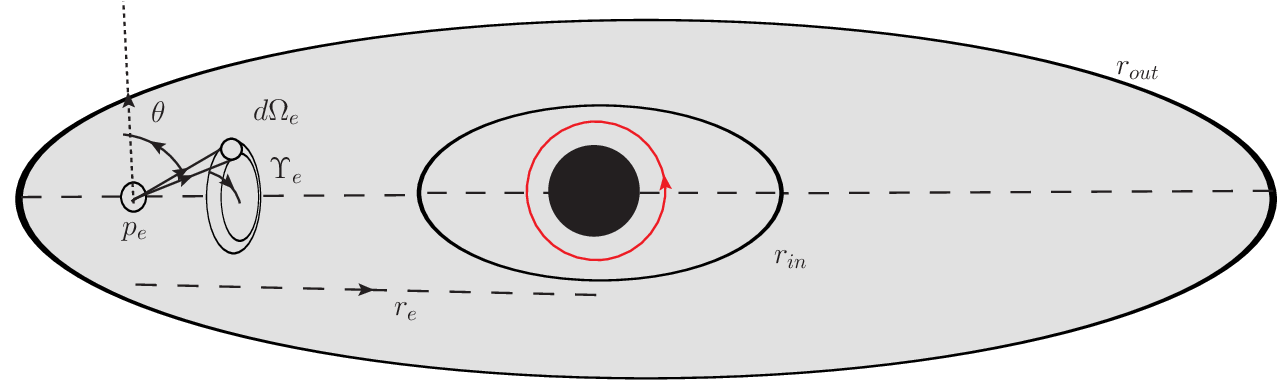}
		\caption{Photons coming from a spherical region in between $r_{in}$ and $r_{out}$ towards the black hole photon sphere which is  shown in red circle.}
		\label{fig:spherical-geometry}
	\end{figure}

	We picture the black hole as being located in the centre of a sphere, where photons are isotropically emitted from each point with a specific emissivity. Here, we provide a ballpark estimate of the number of photons that approach a photon sphere. For this purpose, we will also consider a spherically symmetric spacetime as our model geometry. Let the number of photons with a frequency width of $dw_e$, emitted from $dV_e$, and traveling through an infinitesimal solid angle of $d\Omega_e$, per unit time $\tau_{e}$ as seen from the emission point $p_{e}$ be expressed as
	\begin{equation}
		d^{6}\left(\dfrac{dN}{d\tau_{e}} \right)=J_e(\omega_e, r_e)d\Omega_edV_e d\omega_e
		\label{eqA1}
	\end{equation}
	Here, the volume $V_e$, frequency $\omega_e$, time $\tau_e$, solid angle $\Omega_e$ are all measured in a local inertial frame at $p_{e}$, as shown in \ref{fig:spherical-geometry}. Choosing the point $p_{e}$ as the origin,  let us denote $\Psi_{e}$ as the azimuthal angle in the plane normal to the direction of the black hole from point $p_{e}$ and $\Upsilon_{e}$ as the zenithal angle measured from that direction. With more simplification, we assume that the emission from the point $p_{e}$ is isotropic. With this assumption, we can safely choose $J_{e}$ to be independent of $\Upsilon_{e}$ and $\Psi_{e}$.
	
	We express equation \ref{eqA1} in terms of the impact parameter $b$ of a photon. We use \ref{B6} and $db=\dfrac{r_e}{\sqrt{f(r_e)}}\cos\Upsilon_{e} d\Upsilon_e $ to find the desired result, for a fixed $r_e$. Integrating \ref{eqA1} over the angle $\Psi_e$, we have the number of photons, emitted in the direction of photon sphere, per unit of time as
	\begin{align}\label{Eq:PhNUM}
		d^{5}\left(\dfrac{dN}{d\tau_{e}} \right)=\frac{1}{2}\times 2\pi J_e(\omega_e,r_e)\frac{\sqrt{f(r_e)}}{r_e}\frac{b}{\sqrt{\dfrac{r_e^2}{f(r_e)}-b^2  }}\times db dV_e d\omega_e
	\end{align}
	Due to the fact that only photons with $0\leq \Upsilon_{e}\leq \pi/2$ can approach the photon sphere, the $1/2$ factor is used in \ref{Eq:PhNUM}.
	In a local inertial frame at $p_e$, the time element $d\tau_e$ and volume element $dV_e$ are given by $d\tau_e=\sqrt{f(r_e)}dt$ and $dV_e=\dfrac{r_e^2}{\sqrt{f(r_e)}}\sin\theta dr_e  d\theta d\phi$ in spherically symmetric spacetime (SSS) coordinate.
	The number of photons with impact parameter lying in the range $(b,b+db)$, released from a spherical shell with a width of $dr_{e}$ and  a unit frequency of $\omega_e$ is obtained by integrating \ref{Eq:PhNUM} over $\theta$ and $\phi$, which is given by
	\begin{equation}
		\left(\dfrac{d^{4}N}{ dt d\omega_{e}db dr_{e}}\right)=4\pi^2 J_e(\omega_e,r_e)\frac{br_e\sqrt{f(r_e)}}{\sqrt{(r_e^2/f(r_e))-b^2}} 
	\end{equation}
	It should be noted that $\omega_e$ is the frequency in a local inertial frame at the emission point $p_e$, which is given by
	\begin{equation}
		\omega_e=k^\mu e^{(0)}_\mu\Bigg|_{p_e}=\frac{dt}{d\lambda}\sqrt{f(r_e)}\Bigg|_{p_e}=\frac{E}{\sqrt{f(r_e)}}
	\end{equation}
	where $k^{\mu}=\dfrac{dx^{\mu}}{d\lambda}$ is the tangent vector to the geodesic, $\lambda$ is the affine parameter and $e^{(\alpha)}_{\mu}$ is the local tetrad at $p_{e}$. The frequency measured in a local inertial frame at the photon sphere located at $r=r_{ph}$ is $\omega_c=E/\sqrt{f(r_{ph})}$. 
	As a result, the number of photons reaching the photon sphere per unit time $t$ and unit frequency $\omega_c$ with impact parameter $(b,b+db)$ is given by
	\begin{equation}\label{Eq:PhNUM_b}
		\left( \dfrac{d^{3}N}{dtd\omega_{c}db}\right)=4\pi^2\int_{r_{\rm in}} ^{r_{\rm out}} \dd r_e J_e\left(\frac{\sqrt{f(r_{ph})}\omega_c}{\sqrt{f(r_e)}},r_e \right)\times \frac{br_e\sqrt{f(r_e)}}{\sqrt{(r_e^2/f(r_e))-b^2}} 
	\end{equation}
	where the emission region is in between a spherical region with inner diameter $r_{\rm in}$ and outer diameter $r_{\rm out}$. Since we are concerned about the region near the photon sphere, integrating \ref{Eq:PhNUM_b} in the range $(b_{c},b_{c}(1+\mathfrak{a}\epsilon^{2}))$, we find
	\begin{flalign}\label{Eq:PhNum_NC}
		\dfrac{d^{2}N}{dtd\omega_{c}}\simeq 4\pi^{2}\mathfrak{a} \epsilon^{2} b_{c}^{2}\int_{r_{\rm in}}^{r_{\rm out}} \dd r_{e} J_{e}\left(\dfrac{\sqrt{f(r_{ph})}\omega_{c}}{\sqrt{f(r_{e})}}, r_{e} \right) \times \dfrac{r_{e} \sqrt{f(r_{e})}~ }{\sqrt{(r_{e}^{2}/f(r_{e}))-b_{c}^{2}}} 
	\end{flalign}
	Here, it has been assumed that $r_{e}$ is sufficiently larger than the radius of photon sphere such that $r_{\rm in}^{2}/f(r_{\rm in})>b^{2}$ holds for $b\in (b_{c}, b_{c}(1+\mathfrak{a}\epsilon^{2}))$.
	
	
	\subsection{Sources of photons near galactic black holes}\label{source_ph}
	From recent observations, it is known that in the vicinity of a supermassive black hole electromagnetic plasma exists\cite{EventHorizonTelescope:2019pgp}. Photons with frequencies much larger than the plasma frequency ($\omega\gg \omega_{\rm pl}$) can travel through this plasma. The presence of such high-frequency radiation is a result of a charged particle being accelerated in the Coulomb field of another charge. This phenomenon is known as free-free emission or bremsstrahlung. To fully comprehend this process, a quantum treatment is necessary, as it allows for the production of photons with energies similar to that of the emitting particle. The radiated energy by this process per unit time, per unit frequency, and unit volume is given by\cite{bremmsstrulung}
	\begin{align}
		\left(	\frac{dW}{d\tau_ed\omega_edV_e}\right)=\frac{2^4 \alpha^{3}}{3m_e}\left(\frac{2\pi}{3m_e}\right)^{1/2}T_e ^{-1/2}n_e ^2 e^{-\omega_e/T_e}\bar{g}_{ff}
	\end{align}
	where $T_e$ is the electron temperature, $n_e$ is the electron number density in the plasma and $\bar{g}_{ff}$ is velocity averaged Gaunt factor. We express the emission rate as an approximate classical result times the free emission Gaunt factor $\bar{g}_{ff}$ accounting  for quantum-mechanical Born approximation. $\bar{g}_{ff}$ is a certain function of energy of the electron and of the frequency of emission. But for order magnitude estimation, this factor can be considered to be unity approximately. We have assumed that ion density $n_i$ is equal to $n_e$. For isotropic radiation $J_e^{(N)}$ as defined in \ref{eqA1} leads to
	\begin{align}\label{Eq:Ph_SRC}
		J_e^{(N)}(\omega_e,r_e)=\frac{1}{4\pi \omega_e}\left(\frac{2^4 \alpha^{3}}{3m_e}\right)\left(\frac{2\pi}{3m_e}\right)^{1/2}T_e ^{-1/2}n_e ^2 e^{-\omega_e/T_e}\bar{g}_{ff}
	\end{align}
	In the case of M87*, observations from the Event Horizon Telescope (EHT) have provided evidence for the presence of a hot accretion disk. The EHT images of M87* show a bright ring-like structure, corresponding to the emission from the inner part of the accretion disk. The observed properties of the emission, such as its spectrum and variability, are consistent with expectations from a hot accretion flow. Hot accretion refers to the accretion of gas with high temperatures, typically in the range of millions to billions of Kelvin. The temperature of hot accretion is almost virial, { $T\simeq G M m_{p}/6k_B r \approx10^{12}K (r/M)^{-1}$} with $m_p$ being proton mass\cite{Quataert:2002xn,Narayan:hotaccre}. The study of accretion processes around supermassive black holes like M87* often involves the use of theoretical models to understand the properties and behaviour of the accreting material. One commonly used model is the spherical accretion model, which assumes that the accretion flow onto the black hole is spherically symmetric. It is important to note that while the spherical accretion model provides a useful starting point for understanding the overall behaviour of the accretion flow around M87*, it is a simplification of the actual complex processes that may occur. Then the spherical mass accretion rate can be written as $\dot{M}=4\pi r^2 \rho v_r$ with mass density $\rho$ and radial velocity $v_r$. Assuming constant mass accretion and free falling gas $v_r\propto r^{-1/2}$, we have $\rho\propto (r/M)^{-3/2}$. Then we can assume that the electron temperature and number density of electrons obey the power law:
	\begin{subequations}
		\begin{align}
			T_e&= T_{e,c} \left(\frac{r_e}{r_{ph}}\right)^{-1}\\
			n_e&= n_{e,c} \left(\frac{r_{e}}{r_{ph}}\right)^{-3/2}
		\end{align}
	\end{subequations}
	where $T_{e,c}$ and $n_{e,c}$ are the values at the photon sphere.

	
	\section{Conversion of photons into axions and relative luminosity}\label{Sec:PH_DIM}
	\subsection{Dependence of the conversion probability and conversion factor  on spacetime metric }
	In \ref{Sec:BH_PH}, we have made a discussion on the existence of the photon sphere in a general spherically symmetric spacetime and also derived the photon time-lapse in such  spacetimes. As a further simplification, we choose the general spherically symmetric metric  in the form
	\begin{equation}
		ds^2 = -f(r)dt^2 + \frac{1}{f(r)}dr^2 + r^2(d\theta^2 + \sin^2{\theta}d\phi^2)
	\end{equation}
	where $f(r)$ corresponds to any spherically symmetric black hole geometry. The time that photons stay close to the photon sphere while their impact parameter ($b$) is close to ($b_c$) around the photon sphere ($r_{ph}<r<r_{ph}(1+\delta r)$) is given by
	\begin{equation}
		\T(b) = -\frac{1}{f(r_{ph})}\frac{r_{ph}}{\sqrt{\p}}\ln\left[\frac{2(b-b_{c})}{\p b_{c}}\frac{r_{ph}^2}{\epsilon^2 M^2}\right]
	\end{equation}
	To calculate the number of photons that transform to axions, near the photon sphere, per unit time $t$ and unit frequency $\omega_c$, 	the following equation can be used 
	\begin{align}\label{Eq:Ph_Num}
		\frac{d^{2}N_{\gamma \rightarrow \Phi}}{dt d\omega_c} = \int_{b_c}^{b_c\left(1+\mathfrak{a} \epsilon^2\right)} \dd b~ \frac{1}{2}\left(\dfrac{d^{3}N}{dtd\omega_{c}db}\right)P_{\gamma\rightarrow \Phi}\left(\sqrt{f(r_{ph})}~\T(b)\right)
	\end{align}
	where
	\begin{align}
		P_{\gamma\rightarrow \Phi}\left(\sqrt{f(r_{ph})}~\T(b)\right) = \left(\frac{2\Delta_{\rm M}}{\Delta_{\rm osc}}\right)^{2}\times \sin^{2}\left(-\frac{\Delta_{\rm osc}}{2}\frac{r_{ph}}{\sqrt{\p}\sqrt{f(r_{ph})}}\ln\left[\frac{2(b-b_{c})}{\p b_{c}}\frac{r_{ph}^2}{\epsilon^2 M^2}\right] \right) 
	\end{align}
	The factor of $1/2$ introduced in \ref{Eq:Ph_Num} accounts for the conversion of photons only with polarization parallel to the external magnetic field into axions.
	By integrating over the interval ($b_{c}, b_{c}(1+\mathfrak{a} \epsilon^2)$),  the number of photons that enter the region	 $r_{ph}<r<r_{ph}\left(1+{\epsilon M}/{r_{ph}}\right)$ and escape to infinity can be determined. As no significant change in  the number of photons occurs  with the change in  impact parameter in this scenario, it can be approximated by taking its value at $b=b_c$ out of the integral. Then the \ref{Eq:Ph_Num} can be written as
	\begin{align}
		\frac{d^{2}N_{\gamma \rightarrow \phi}}{dt d\omega_c} &\simeq \frac{1}{2} \hspace{0.1cm}\left(\dfrac{d^{3}N}{dtd\omega_{c}db}\right)\Bigg|_{b=b_c}\times \int_{b_c}^{b_c\left(1+\mathfrak{a} \epsilon^2\right)} \dd b\ P_{\gamma\rightarrow \Phi}\left(\sqrt{f(r_{ph})}\T(b)\right)\\
		&=\left.\dfrac{1}{2}\left(\dfrac{d^{3}N}{dtd\omega_{c}db}\right)\right|_{b=b_{c}} \times
		\left(\dfrac{2\Delta_{\rm M}}{\Delta_{\rm osc}}\right)^{2}\times \int_{b_{c}}^{b_{c} \left(1+\mathfrak{a}  \epsilon ^2\right)}  \sin^2\left(\frac{\Delta_{\text{osc}}}{2} ~\T(b)\sqrt{f(r_{ph})}\right) \dd b \label{Eq:PhNUM_sin}
	\end{align}
	The associated integral can be simplified as
	\begin{flalign}
		I&=\int_{b_{c}}^{b_{c} \left(1+\mathfrak{a}  \epsilon ^2\right)}  \sin^2\left(\frac{\Delta_{\text{osc}}}{2} ~\T(b)\sqrt{f(r_{ph})}\right) \dd b\\
		&=\dfrac{1}{2}\mathfrak{a} b_{c} \epsilon^{2}\left[1 -  \dfrac{\sqrt{\p} \sqrt{f(r_{ph})}}{\Delta_{\rm osc}\ r_{ph}}\int^{\infty}_{{0}}{\cos x}\ \exp\left(-\dfrac{\sqrt{\p}\sqrt{f(r_{ph})}}{\Delta_{\rm osc}r_{ph}}\ x \right) \dd x \right]\nonumber\\
		&=\dfrac{1}{2}\left( \dfrac{1}{1+\mathcal{X}^{2}}\right)\mathfrak{a} b_{c}\epsilon^{2} 
	\end{flalign}
	where we have defined the dimensionless quantity $\mathcal{X}$ as
	\begin{flalign}
		\mathcal{X}=\dfrac{\sqrt{\p}\sqrt{f(r_{ph})}}{\Delta_{\rm osc} r_{ph}}
	\end{flalign}
	and we have used the relation $\dfrac{2\mathfrak{a} r_{ph}^{2}}{\p M^{2}}=1$.
	
	The fraction of photons entering the region near the photon sphere that are converted into axions is given as
	\begin{align}\label{conversion factor}
		\frac{d^{2}N_{\gamma \rightarrow \phi}}{dt d\omega_c}\bigg/\dfrac{d^{2}N}{dtd\omega_{c}}\simeq \dfrac{1}{4} \left(\frac{2\Delta_{\rm M}}{\Delta_{\rm osc}}\right)^{2}  \left(\dfrac{1}{1+\mathcal{X}^{2}} \right) 
	\end{align}
	For the efficient conversion, we have $\Delta_{\rm osc}\simeq 2\Delta_{\rm M}$. For such a scenario, the conversion factor (CF) can be given by
	\begin{flalign}
		CF=	\frac{dN_{\gamma \rightarrow \Phi}}{dt d\omega_c}\bigg/\dfrac{d^{2}N}{dtd\omega_{c}} \Bigg|_{\text{ efficient conversion}}&\simeq \dfrac{1}{4}\left(\dfrac{1}{1+\mathcal{X}^{2}} \right)\\
		&=\dfrac{1}{4}\left[\dfrac{\left(\dfrac{r_{ph}}{M} \right)^{2} \left(2 M\Delta_{\rm M}\right)^{2}}{\left(\dfrac{r_{ph}}{M} \right)^{2} \left(2 M\Delta_{\rm M}\right)^{2}+{\p f(r_{ph})}} \right] 
	\end{flalign}
	Now, inserting \ref{Eq:PhNum_NC} into the \ref{Eq:PhNUM_sin}, we have
	\begin{flalign}\label{Eq:Axion_Num}
		\dfrac{d^{2}N_{\gamma\to\Phi}}{dtd\omega_{c}}&\simeq {\mathfrak{a} \pi^{2}\epsilon^{2}}b_{c}\left(\dfrac{2\Delta_{\rm M}}{\Delta_{\rm osc}} \right)^{2}\left[\dfrac{\left(\Delta_{\rm osc}r_{ph}\right)^{2}}{\left(\Delta_{\rm osc}r_{ph}\right)^{2}+\p f(r_{ph})} \right] \nonumber\\
		&\hspace{4cm}\times \int_{r_{\rm in}} ^{r_{\rm out}} \dd r_e  J_e\left(\frac{\sqrt{f(r_{ph})}\omega_c}{\sqrt{f(r_e)}},r_e \right)\left[ \frac{br_e\sqrt{f(r_e)}}{\sqrt{(r_e^2/f(r_e))-b^2}} \right]
	\end{flalign}
	The above equation i.e \ref{Eq:Axion_Num} enable us to find the number of axions created via the conversion of the photons to axions. 
	
	\subsection{Spherically symmetric spacetime in brane-world scenario }
	Here, we will discuss the spherically symmetric spacetime in the presence of extra dimension i.e in  the brane-world scenario. The effective gravitational field equations on the brane can be solved for a spherically symmetric scenario in the vacuum by projecting the bulk gravitational field equations onto the brane hypersurface. This method of getting the gravitational field equations on the brane is feasible because it avoids the need for extensive knowledge of the bulk geometry, which is presumptively lacking for any observer of the brane. If the bulk gravitational field equations are the bulk Einstein equations, then the effective gravitational field equations on the brane can be obtained by projecting the bulk Einstein tensor onto the brane hypersurface using the Gauss, Codazzi, and Mainardi equations\cite{Shiromizu:1999wj,Harko:2004ui,Chakraborty:2014xla}. The  effective field equations of vacuum brane consist of the following terms
	\begin{itemize}
		\item The four dimensional Einstein tensor $G_{\mu\nu}$, not relating to the additional spatial dimension.
		\item Projection of the bulk Weyl tensor $W_{ABCD}$, onto the brane hypersurface, which yield  $E_{\mu\nu}=W_{ABCD}e^{A}_{\mu}e^{C}_{\nu}n^{B}n^{D}$, where $e^{A}_{\mu}$ are the projectors and $n_{A}$ are the normals to the brane hypersurface.
	\end{itemize}
	We would not address the additional contributions from the matter sector in the case of non-vacuum brane at this time. In the effective gravitational field equations, the term $E_{\mu\nu}$, which is derived from the bulk Weyl tensor, incorporates the fact that vacuum branes have an additional spatial dimension. Due to its relationship to the Weyl tensor and the symmetries of that tensor, it follows that the tensor $E_{\mu\nu}$ is traceless and so resembles the energy-momentum tensor of a Maxwell field with a crucially important overall negative value\cite{Aliev:2005bi}. As a result of the effective gravitational field equations on the brane, the static and spherically symmetric vacuum brane spacetime has already been studied in \cite{Dadhich:2000am}.
	Due to the fact that the tensor $E_{\mu\nu}$ shares the same characteristics as the energy-momentum tensor of the Maxwell field, the resulting static spherically symmetric solution will resemble \RN spacetime, with the accompanying line element having the following form
	\begin{flalign}\label{metric}
		ds^{2}=-f(r)\ dt^{2}+\dfrac{1}{f(r)} dr^{2}+ r^{2}\left(d\theta^{2}+\sin^{2}\theta d\phi^{2} \right)
	\end{flalign}
	with 
	\begin{flalign}\label{spherical_geomtry}
		f(r)=\left(1-\dfrac{2M}{r}+\dfrac{\beta M^{2} }{r^{2}} \right)
	\end{flalign}
	Here $M$ is the mass parameter of the black hole. The dimensionless  parameter $\beta$, which appears in \ref{spherical_geomtry}, is inherited from the presence of higher dimensions and therefore can take on positive and negative values. This is in stark contrast to the Maxwell field, where the corresponding contribution to the metric components is made through $Q^{2}$, which is strictly positive regardless of the sign of the electric charge $Q$. Unlike black holes with an electromagnetic field, where  \ref{metric} resembles a \RN black hole with an event horizon and a Cauchy horizon for positive values of $\beta$, however, the scenario for negative values of $\beta$ has no parallel in general relativity and so offers a real indicative of the extra spatial dimensions\cite{Aliev:2005bi}. We will look for schwarzchild metric when $\beta$ is zero as well as for the black holes with positive $\beta$ and negative $\beta$ values as special cases.
	
\begin{figure}[h]%
		\centering
		\subfloat[The variation of the conversion factor(CF) with change in $M\Delta_{M}$ has been shown here for different values of the extra-dimensional parameter $\beta$. The conversion factor reaches its maximum value when $M\Delta_{M}$ approaches  unity irrespective of the value of the extra-dimensional parameter $\beta$.  ]{{\includegraphics[width=7.2cm,height=5cm]{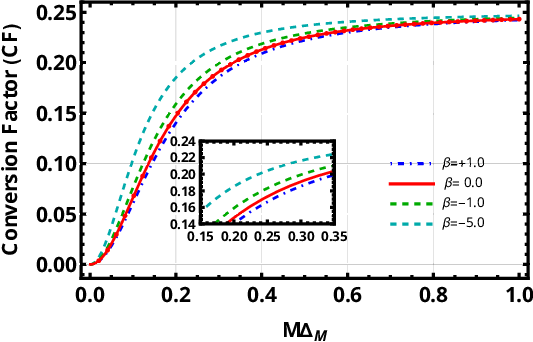} } \label{Fig:CF_1}}
		\qquad
		\subfloat[Here, for various values of $M\Delta_{M}$, the variation of the conversion factor (CF) with change in the extra-dimensional parameter $\beta$ is displayed. At $\beta=27/32$, conversion factor reaches its minimum value for any fixed value of $M\Delta_{M}.$]{{\includegraphics[width=7cm,height=5cm]{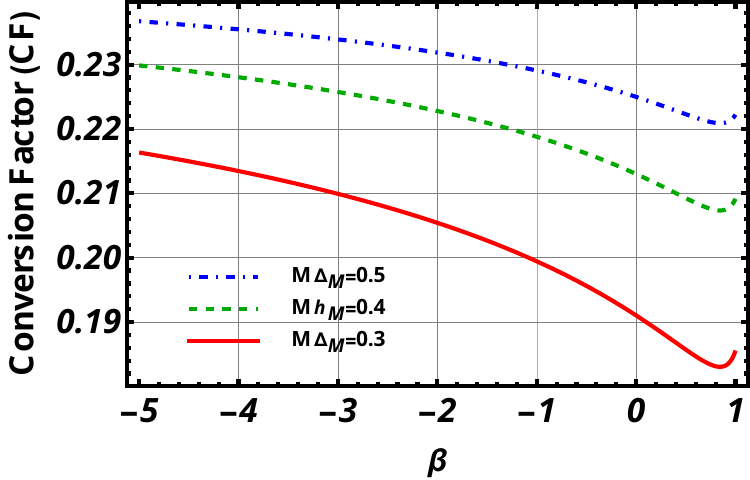} } \label{Fig:CF_2}}
		\qquad
		\subfloat[This contour plot shows higher values of $M\Delta_{M}$ and more negative values of extra-dimensional parameter $\beta$ make the conversion more effective. ]{{\includegraphics[width=8cm,height=6cm]{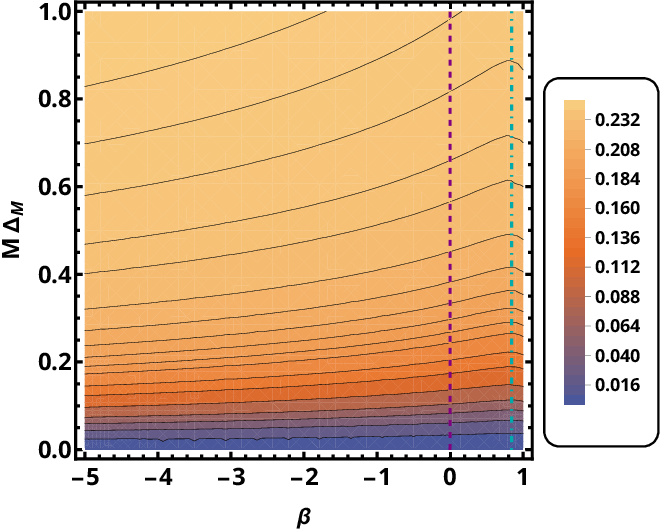} } \label{Fig:CF_3}}
		\caption{For efficient conversion i.e when $2\Delta_{M}=\Delta_{\rm osc}$, the variation of conversion factor(CF) has been demonstrated with variation of $M\Delta_{M}$ and the extra-dimensional parameter $\beta$. }
		\label{Fig:CF}%
	\end{figure}

	In \ref{Fig:CF} we have displayed the variation of the conversion factor, in the effective conversion regime, with the variation  of  $M\Delta_{M}$ and the extra-dimensional parameter $\beta$. In such regime, the conversion factor(CF) increases with the increase in axion-photon couplings for obvious reasons.  If one choose for efficient conversion, the conversion factor(CF) reaches its lowest value for $\beta=\dfrac{27}{32}$ as shown in \ref{Fig:CF_2}. On the other hand, when $\beta$ takes on negative values, the dimming intensifies, revealing a distinct indication of the presence of an additional dimension. For a specific black hole which has a specific mass and magnetic field associated with it, the variation of conversion factor with the variation of  dimensionless quantity $M\Delta_{M}$  is of immense importance. This variation, as shown in \ref{Fig:CF_1}, suggests that the increasing value of $M\Delta_{M}$ or equivalently  the increasing value of axion-photon coupling constant leads to  higher conversion rate of photons into axions. Such kind of variation of conversion factor with change in coupling constant holds for any value of extra-dimensional parameter $\beta$.

	\subsection{Dimming of the photon ring luminosity and observation of the axions}\label{Fading of the ring}
	\subsubsection{Dimming of the photon sphere}
	Once the conversion factor is known to us, we only need a count of the photon numbers that approached the photon sphere and escape to infinity to find the number of converted photons and newly generated axions. Using \ref{Eq:PhNum_NC} and \ref{Eq:Ph_SRC}, we have the photon count as
	\begin{flalign}
		\dfrac{d^{2}N}{dtd\omega_{c}}
		&= L_{\omega}^{0}\omega_{\rm obs}^{-1}\ 2 \mathfrak{a} \left(\dfrac{b_{c}}{M}\right)^{2} \left(\dfrac{r_{ph}}{M}\right)\times\int_{x_{\rm in}}^{x_{\rm out}} \dd x_{e} \left[\dfrac{x_{e}^{-3/2}f(x_{e})}{\sqrt{x_{e}^{2}/f(x_{e})-(\frac{b_{c}}{r_{ph}})^2}}\right] e^{-\dfrac{\omega_{\rm obs}}{T_{e,c}}\dfrac{x_{e}}{\sqrt{f(x_{e})}}}\label{Eq:LUM}
	\end{flalign}
	where we have defined the variable $x$ as $x={r}/{r_{ph}}$ and also $L_{\omega}^{0}$ is defined as
	\begin{flalign}
		L_{\omega}^{0}&=2  \pi^{2}\epsilon^{2}M^{3}\left(\dfrac{4\alpha^{3}}{3\pi m_{e}}\right)\left(\dfrac{2\pi}{3m_{e}} \right)^{1/2} T_{e,c}^{-1/2}n_{e,c}^{2}\ \bar{g}_{ff} \nonumber\\
		&= 6.48\times 10^{27} \text{erg}\cdot \text{sec}^{-1}\cdot \text{KeV}^{-1} \epsilon^2 \left(\frac{M}{6.2\times10^9 M_{\odot}}\right)^3 \left(\frac{T_{e,c}}{10^{11} K}\right)^{-1/2} 
		\left(\frac{n_{e,c}}{10^4 \text{cm}^{-3}}\right)^2 \Bar{g}_{ff}
	\end{flalign}
	The observed frequency $\omega_{\rm obs}$  is related to that frequency $\omega_c$ at the photon sphere as $\omega_{\rm obs}=\sqrt{f(r_{ph})}\omega_c$ due to gravitational redshift. The distance to M87$^{*}$, denoted as D, is approximately 16.8 Mpc. The cosmological red-shift of M87$^{*}$ is estimated to be around  $z\approx \frac{H_0 D}{c}$ $ \approx 0.004$, where H$_0$ represents the present value of the Hubble parameter (taken to be 71 km/s/Mpc) and c is the speed of light.
	In the local universe, galaxies, including Messier galaxy, typically exhibit peculiar velocities on the order of a few hundred kilometers per second. However, when considering the dominant gravitational redshift, we neglect small effects such as peculiar velocities and cosmic expansion.
	
\begin{figure}[h!]%
		\centering
		\subfloat[Dimmed and undimmed relative luminosity spectrum of photons has been depicted assuming the axion mass to be $m_{\Phi}=1 {\rm neV}$ for different values of extra-dimensional parameter $\beta$.]{{ \includegraphics[scale=0.6]{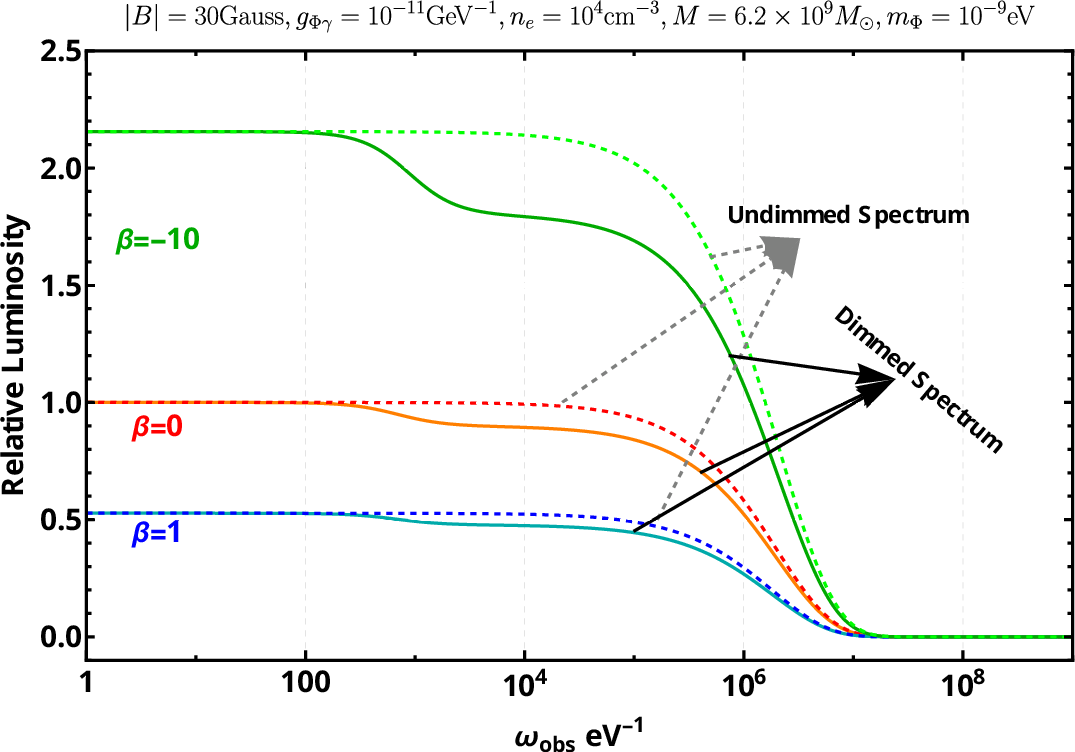} } \label{Fig:RL_1}}
		\qquad
		\subfloat[Dimmed and undimmed relative luminosity spectrum of photons has been depicted assuming the axion mass to be $m_{\Phi}=10 {\rm neV}$ for different values of extra-dimensional parameter $\beta$. ]{{ \includegraphics[scale=0.6]{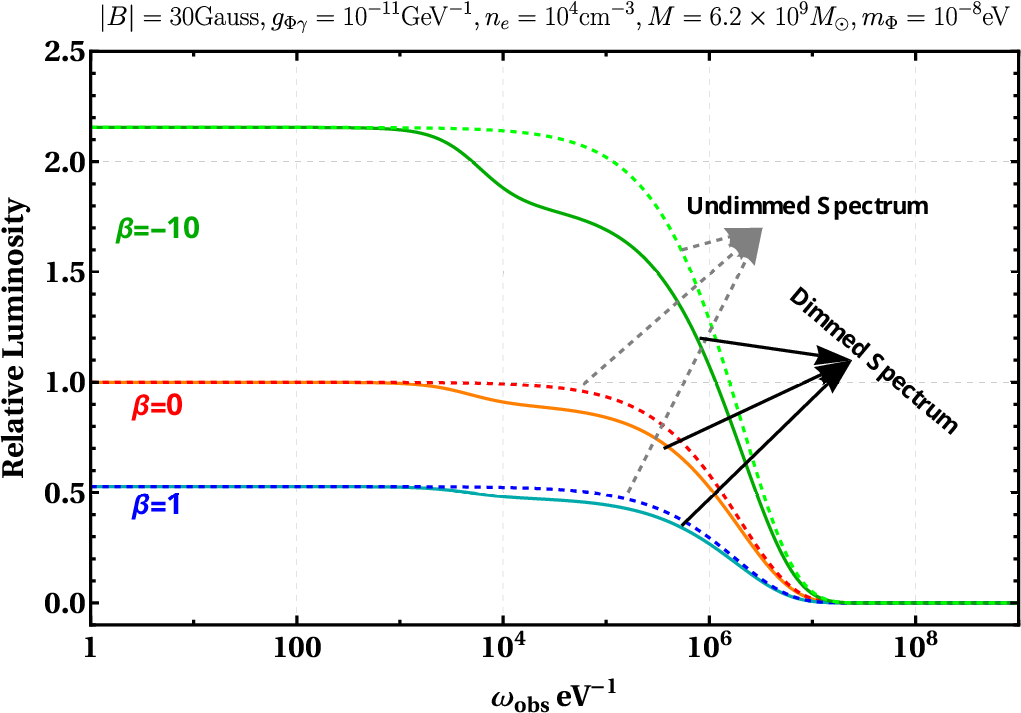} } \label{Fig:RL_2}}
		\caption{ Relative Luminosity of photons has been displayed before (with dashed lines) and after (with solid lines) the photon-axion conversion. Here we have assumed the magnetic field associated with the black hole to be $\B=30\ {\rm Gauss}$, the axion-photon coupling to be $ g_{\Phi\gamma}=10^{-11}\ {\rm GeV^{-1}}$, electron number density to be $ n_{e}=10^{4}\ {\rm cm^{-3}}$ and the mass of the black hole to be $ M=6.2\times 10^{9}\ M_{\odot}$. Plots have been displayed for two different mass values of axions. }
		\label{Fig:RL}%
	\end{figure}

	We define the following quantity to be the \emph{relative luminosity} (RL) of  photons before any conversion as
	\begin{flalign}
		\dfrac{L_{\omega}^\gamma}{L_{\omega}^{0}} 
		&\equiv \dfrac{\omega_{obs}}{L_{\omega}^{0}} \left( \dfrac{d^{2}N}{dt d\omega_{c}} \right)= 2 \mathfrak{a} \left(\dfrac{b_{c}}{M}\right)^{2} \left(\dfrac{r_{ph}}{M}\right)\times\int_{x_{\rm in}}^{x_{\rm out}} \dd x_{e} \left[\dfrac{x_{e}^{-3/2}f(x_{e})}{\sqrt{x_{e}^{2}/f(x_{e})-\left(\frac{b_{c}}{r_{ph}}\right)^2}}\right] e^{-\dfrac{\omega_{\rm obs}}{T_{e,c}}\dfrac{x_{e}}{\sqrt{f(x_{e})}}}
	\end{flalign}	
	The relative luminosity for the axions as produced during photon-axion conversion is given by
	\begin{flalign}
		\dfrac{L_{\omega}^{\gamma\rightarrow\Phi}}{L_{\omega}^{0}} 
		&=  \dfrac{L_{\omega}^\gamma}{L_{\omega}^{0}}\frac{1}{4}\left(\frac{2\Delta_{\rm M}}{\Delta_{\rm osc}}\right)^{2}  \left(\dfrac{1}{1+\mathcal{X}^{2}} \right)
	\end{flalign}
	The spectral variation of the relative luminosity(RL) has been depicted in \ref{Fig:RL} for various values of the extra-dimensional parameter $\beta$. The spectrum has been normalised with respect to the undistorted relative luminosity for Schwarzschild black hole in very low frequency. The dashed lines indicate the RL if there is no conversion of the photons, approaching the photon sphere. While the solid lines show that the spectrum is attenuated as a result of the fraction of photons becoming axions. The effect of the attenuation of the spectrum from its undistorted part becomes more visible when the parameter $\beta$ is increased in its negative direction. Moreover, by comparing \ref{Fig:RL_1} and \ref{Fig:RL_2}, one can infer that the attenuation point of the spectrum also depends on the mass value of the  axion. The rapid downtrend of the spectrum occurs due to the presence of the exponential factor in \ref{Eq:LUM} which in turn appeared through the \ref{Eq:Ph_SRC}. Because there are so few high-temperature electrons that release photons with such high frequencies in the electromagnetic plasma, there is a suppression of these photon luminosities.
	The exponential factor in \ref{Eq:Ph_SRC} can be put to unity if the following condition holds
	\begin{flalign}
		\dfrac{\omega_{e}}{T_{e}}=\dfrac{\omega_{e}}{T_{e,c}}\left(\dfrac{r_{e}}{r_{ph}}\right)=\dfrac{\omega_{c}}{T_{e,c}}\sqrt{\dfrac{f(r_{c})}{f(r_{e})}}\left(\dfrac{r_{e}}{r_{ph}} \right) \ll 1
	\end{flalign}
	Under such conditions, the RL reaches its maximum value and an almost flat part of the spectrum is obtained.  
	
\begin{figure}[h!]%
		\centering
		\subfloat[Relative luminosity spectrum of axions for axion mass $m_{\Phi}=1\ {\rm  neV}$ with different values of extra-dimensional parameter.]{{ \includegraphics[scale=0.6]{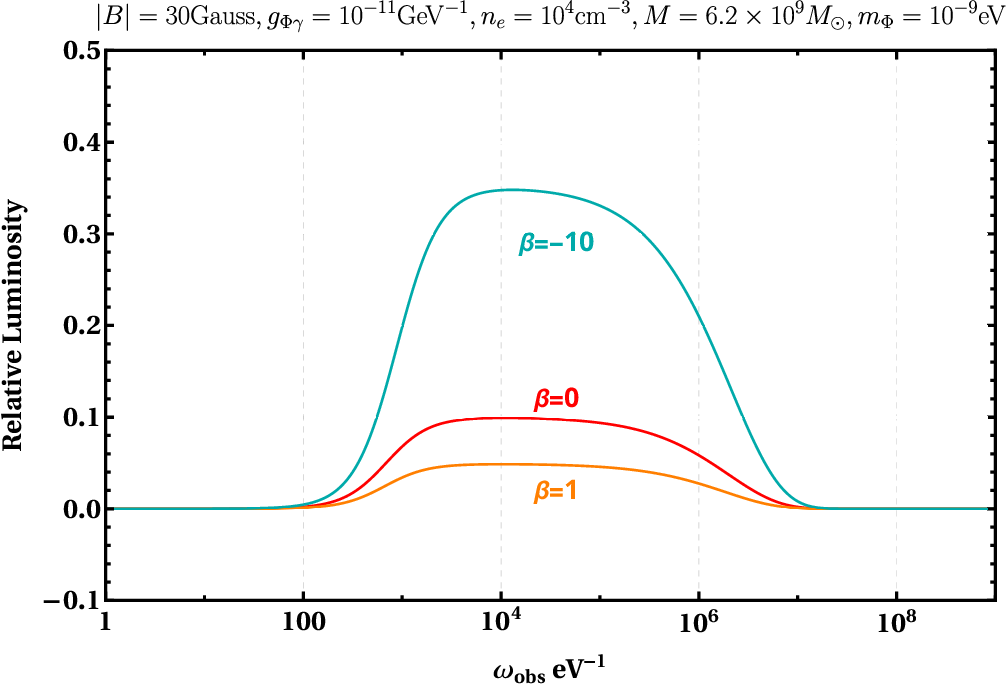} } \label{Fig:RL_Axion_1}}
		\qquad
		\subfloat[Relative luminosity spectrum of axions for axion mass $m_{\Phi}=10\ {\rm  neV}$ with different values of extra-dimensional parameter. ]{{ \includegraphics[scale=0.6]{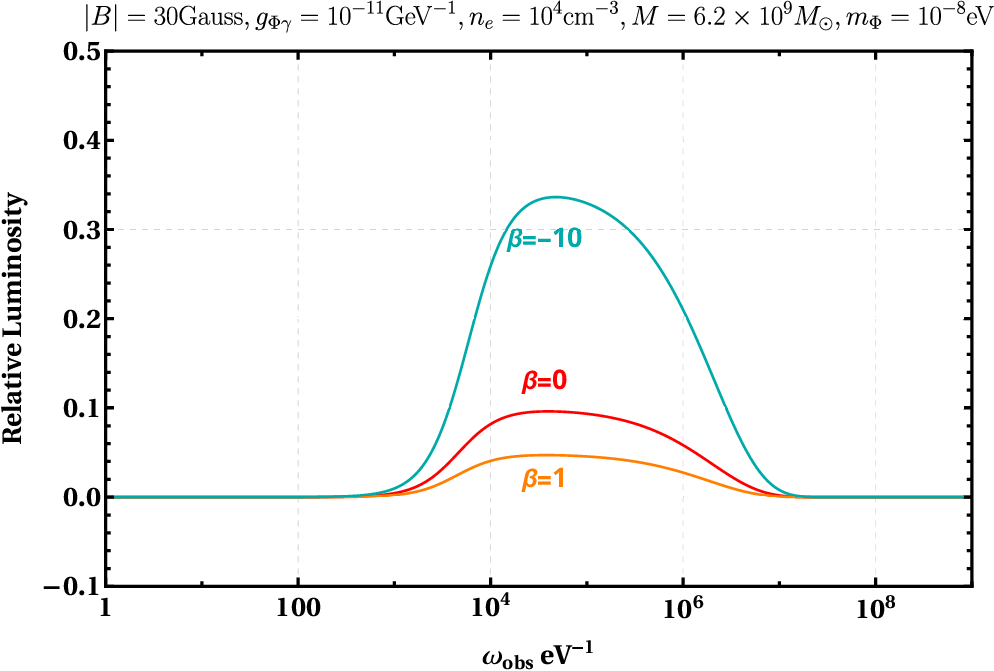} } \label{Fig:RL_Axion_2}}
		\caption{Relative luminosity spectrum of axion has been plotted. Here we have assumed the magnetic field associated with the black hole to be $\B=30\ {\rm Gauss}$, the axion-photon coupling to be $ g_{\Phi\gamma}=10^{-11}\ {\rm GeV^{-1}}$, electron number density to be $ n_{e}=10^{4}\ {\rm cm^{-3}}$ and the mass of the black hole to be $ M=6.2\times 10^{9}\ M_{\odot}$. Plots have been displayed for two different mass values of axions.}
		\label{Fig:RL_Axion}%
	\end{figure}

	In \ref{Fig:RL_Axion}, the relative luminosity spectrum of axion has been depicted. The contributing axions are created from the photons near the photon sphere. From the spectrum, it can again be inferred that the more negative value of the extra-dimensional parameter $\beta$  induces the higher production rate of the axions. Moreover, one can also notice the broadening of the axion spectrum as the mass of the axion decreases. This has explicitly  also been shown in \ref{fine_tuning_axion_mass} and discussed in corresponding subsection.

	\newpage
	\subsubsection{Spectral flux of photons}
	The spectral flux coming from the nearby region of the photon sphere incorporating the photon-to-axion conversion is approximately estimated as\\
	\begin{align}\label{Dimmed_flux}
		F_{\mathfrak{D},\omega} & \simeq \frac{1}{4\pi D^2} \left(\dfrac{d^{2}N}{dtd\omega_{c}}-\frac{d^2 N_{\gamma \rightarrow \Phi}}{dt d\omega_0}\right)\\
		& = 1.2\times 10^{-15} \text{cm}^{-2}\cdot \text{sec}^{-1}\cdot \text{KeV}^{-1} \left(1-\frac{1}{4}\left(\frac{\Delta_{\rm M}}{\Delta_{\rm osc}/2}\right)^{2}  \left[\dfrac{\left(\Delta_{osc}r_{ph}\right)^{2}}{ \left(\Delta_{osc}r_{ph}\right)^{2}+{\p f(r_{ph})}} \right]\right) \notag \\
		& \quad \times 2 \mathfrak{a} \left(\dfrac{b_{c}}{M}\right)^{2} \left(\dfrac{r_{ph}}{M}\right)\times\int_{x_{\rm in}}^{x_{\rm out}} \dd x_{e} \left[\dfrac{x_{e}^{-3/2}f(x_{e})}{\sqrt{x_{e}^{2}/f(x_{e})-(\frac{b_{c}}{r_{ph}})^2}}\right] e^{-\dfrac{\omega_{\rm obs}}{T_{e,c}}\dfrac{x_{e}}{\sqrt{f(x_{e})}}}\nonumber\\ 
		& \quad \times \left(\frac{16.8 Mpc}{D}\right)^2 \left(\frac{M}{6.2\times10^9 M_{\odot}}\right)^3 \left(\frac{T_{e,c}}{10^{11} K}\right)^{-1/2}  \left(\frac{n_{e,c}}{10^4 \text{cm}^{-3}}\right)^2 \left(\frac{keV}{\omega_{obs}}\right)\epsilon^{2}\Bar{g}_{ff}
	\end{align}
	For the supermassive black hole M87*, located at a distance of D = 16.8 Mpc, the conversion reaches its maximum in the frequency range from  nearly few eV to MeV, depending upon the  axion mass for a fixed value of the parameter $\beta$. The luminosity and fluxes of photons in this range vary depending on the extra-dimensional parameter $\beta$ of the black hole  as shown in \ref{Fig:RL}. From the plot, in the case of the Schwarzchild blackhole (i.e. for $\beta=0$), the dimming fraction is about $10\%$. With this fraction and setting the parameters to approximate values, such as $\epsilon \sim \bar{g}_{ff} \sim 1$, the flux is estimated to be  $F_{\mathfrak{D},\omega} \sim 6.28 \times 10^{-15}\times(1-\text{conversion factor}) \text{ cm}^{-2} \cdot \text{sec}^{-1} \cdot \text{KeV}^{-1}$ for $\omega_{obs}$ around the keV range. In comparison, for the black hole SgR A*, although it is closer at a distance of D = 8 Kpcs, its mass is three orders of magnitude smaller (approximately $4 \times 10^{6} M_{\odot}$). Since the flux depends on the mass and distance according to ${M^3}/{D^2}$, the flux for SgR A* is much smaller compared to that of M87* and it is difficult to get a flux from the black hole centred in the Milky Way.
	
	\subsubsection{Observation limit on axion mass}
\begin{figure}[h!]
		\centering
		\includegraphics[width=0.7\textwidth]{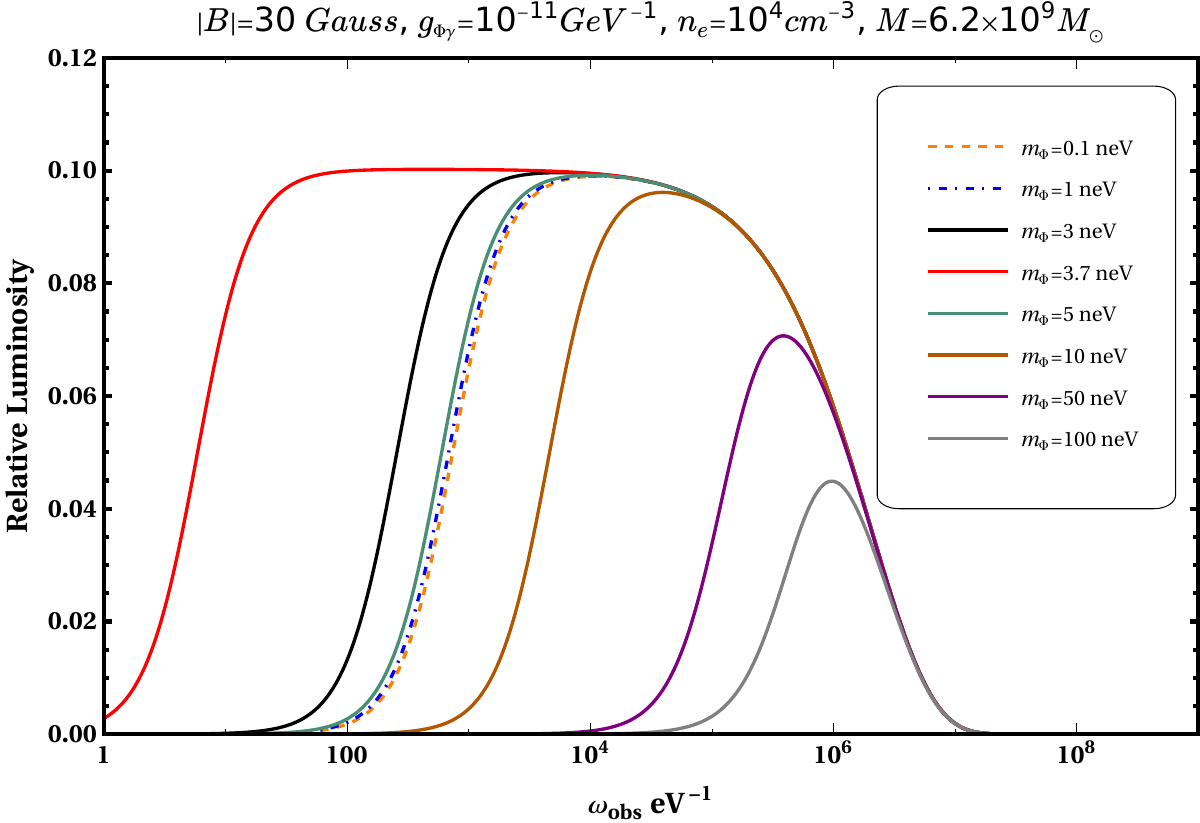}
		\caption{The plot describes the expected spectral luminosity of axions arising from photons in proximity to the photon sphere, considering different axion masses within the framework of a Schwarzschild black hole.}
		\label{fine_tuning_axion_mass}
	\end{figure}

	Referring to  \ref{fine_tuning_axion_mass}, a narrow range of converted axion masses can be demonstrated. When the mass of $m_{\Phi}$ exceeds 100 neV, the dimmed spectra align with the spectra of source photons, resulting in minimal observable impact. Consequently, an upper bound on the mass is imposed, with $m_{\Phi}$ being less than or approximately 100 neV, for effective photon-axion conversion near the photon sphere. Furthermore, if the axion mass surpasses $\sim$ 1 neV, the spectra merge, indicating saturation in mass. In the figure, we observe an intriguing reversal of spectral behavior between 3.7 neV and 1 neV. This is due to specific parameter values where a resonance occurs in photon-axion conversion at an axion mass of 3.7 neV. So the parameter space with finite photon-axion conversion which can give maximum observable opportunity falls in the axion mass range $\approx$ ($1-100$ )neV.
	\begin{figure}
		\centering
		\includegraphics[width=0.8\textwidth,height=0.6\textwidth]{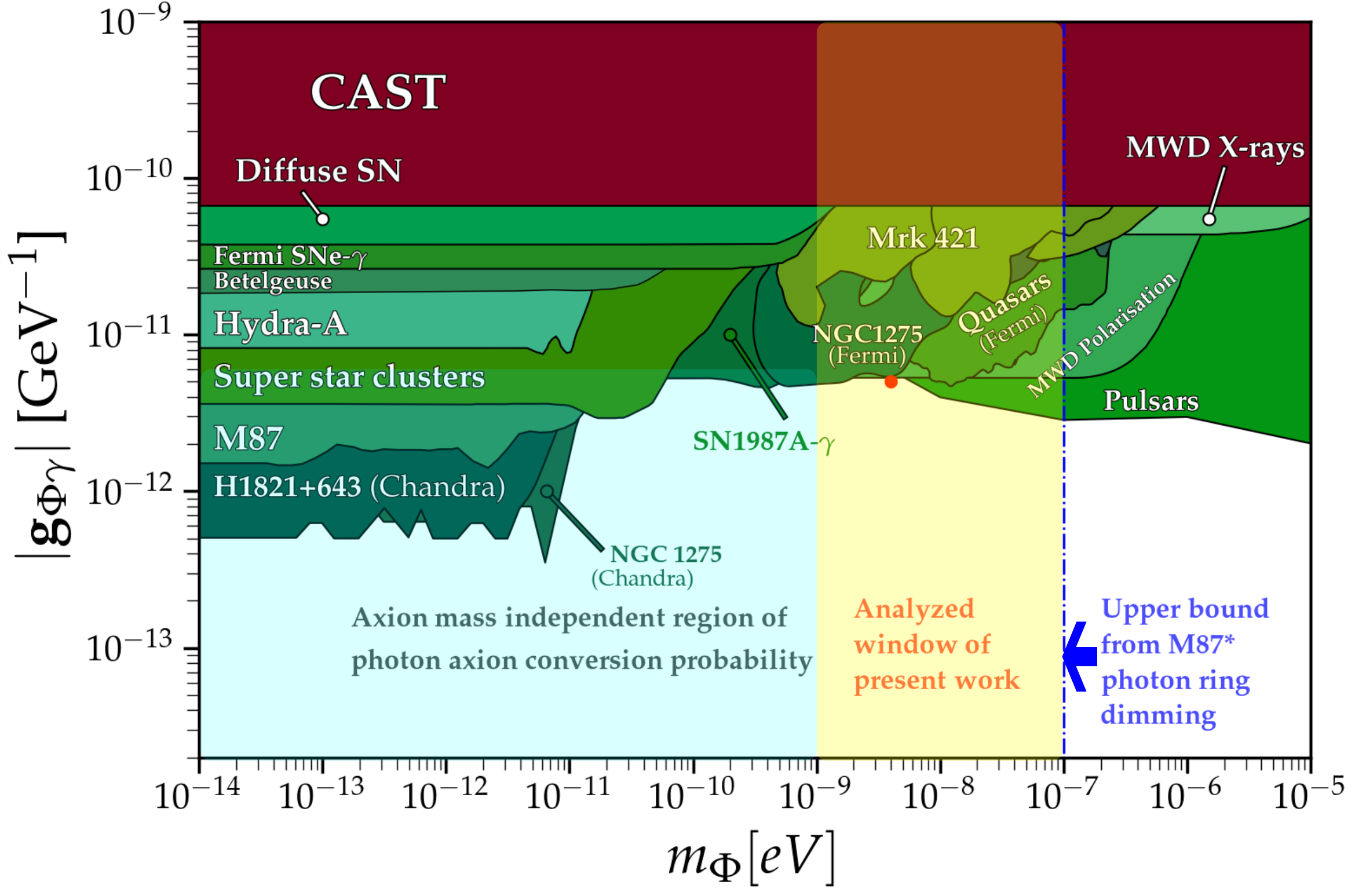}
		\caption{The plot describes the low mass astrophysical bounds on coupling and mass of axion-photon interactions from different observations\cite{AxionLimits}{} . The region in light-yellow demonstrates the window for which the dimming of the photon ring can be expected to be observed. For axion mass $m_{\Phi}\lesssim 10^{-9}$ there is no significant change in photon-axion conversion probability due to a change in axion mass. This mass range has been shown in sky blue region of $g_{\Phi\gamma}-m_{\Phi}$ plane .}
		\label{mass_coupling bounds}
	\end{figure}
	The  plot as shown in \ref{mass_coupling bounds} gives the bounds on coupling and mass of axions from different astrophysical observations. In the present study we have found that there is upper bound on axion mass at $m_{\Phi}=10^{-7} eV$ as demonstrated from the spectra of \ref{Fig:RL}. For axion mass lower than 1neV, the spectra merges as seen from \ref{fine_tuning_axion_mass} . This is the reason why the likelihood of conversion remains unaffected by mass throughout the entire region, which is denoted as sky blue in the figure for  $m_{\Phi}\lesssim$ 1neV. The displayed window having axion mass in the range $10^{-9}$ eV - $10^{-7}$ eV , showcases a valuable parameter space on coupling and mass for the M87* supermassive black hole. In our calculations, we have utilized the coupling constant of $g_{\Phi\gamma}=10^{-11}GeV^{-1}$, which nearly falls within the permissible region on the $g_{\Phi\gamma}-m_{\Phi}$ plane. To be precise, in the plot, we have highlighted a specific data point with a red dot, where the coupling constant and axion mass are approximately $5\times 10^{-12} GeV^{-1}$ and 4 neV, respectively. At this point, the photon-axion conversion probability reaches approximately $3.6\%$ when considering a Schwarzschild black hole( $\beta=0$).
	
	\subsubsection{Required resolution of the black hole image}
	The Chandra X Observatory has achieved an angular resolution of 1 arc second for observing objects beyond the event horizon scale\cite{Uttley:2019ngm} which is more specifically the photons that originated through bremsstrahlung in nearby plasma well outside the photon sphere. The total luminosity of photons emitting from a region of space in between $r_{ISCO}<r_e<\mathfrak{R}$ with energies in keV-MeV band is  approximately given by,
	\begin{align}
		L_{total}&=4\pi \int_{r_{\rm ISCO}}^\mathfrak{R} dr_e r_e {^2} \frac{dW}{d\tau_e d\omega_e dV_e}\\
		&\sim \frac{16\pi}{\epsilon^2}\left(\frac{r_{ph}}{M}\right)^3 \Bigg[\left(\frac{\mathfrak{R}}{r_{ph}}\right)^{1/2} - \left(\frac{r_{ISCO}}{r_{ph}}\right)^{1/2}\Bigg] L_{\omega}^0,
	\end{align}
	\begin{figure}[h!]
		\centering
		\includegraphics[width=0.8\textwidth]{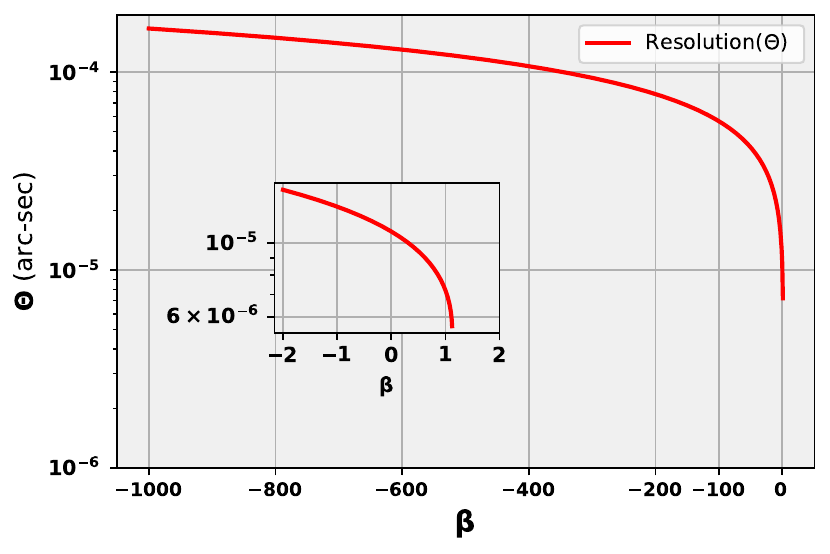}
		\caption{The red plot illustrates the relationship between the charge parameter and the required resolution for observing the dimming effect in a spherically symmetric black hole. Specifically, it demonstrates how the necessary resolution changes as the charge parameter varies. In the context of a Schwarzschild black hole, the point where the plot intersects the $\beta = 0$ line symbolizes the minimum resolution that is required. Inset figure shows the variation of resolution for small $\beta$ values.}
		\label{resolution_vs_beta}
	\end{figure}
	where $r_{ISCO}$ stands for the radius of the  Innermost Stable Circular Orbit. So when observing objects at a distant scale where the radius is much larger than the innermost stable circular orbit ($\mathfrak{R} \gg r_{ISCO}$), the majority of the total luminosity is accounted for, resulting in an insignificant dimming effect caused by the photon-axion conversion. However, when the distance scale is smaller than approximately 3 times the mass of the object ($\mathfrak{R} \lesssim 3M$) i.e. size of photon sphere of a Schwarzchild black hole, this scenario changes. In such cases, the resolution will be 
	\begin{flalign}
		\theta =\frac{\mathfrak{R}}{D}\Bigg|_{\beta=0}\lesssim 1.09 \times 10^{-5} \text{ arc-sec} \left(\frac{M}{6.2\times10^{9} M_\odot} \right)\left(\frac{\text{16.8 Mpc}}{D}\right)
	\end{flalign}
	For the black holes having  spherical symmetry as mentioned in \ref{spherical_geomtry}  , the resolution  will vary with the  $\beta$ parameter as indicated in \ref{resolution_vs_beta}. The photons originating from a thermal bremsstrahlung process are depicted as undimmed spectra in \ref{Fig:RL}, exhibiting a resolution akin to that of the Chandra observatory, approximately 1 arc second. However, to effectively study the region near the photon sphere, it becomes imperative to enhance the angular resolution by at least a factor of $10^5$ compared to Chandra's resolution. Notably, the Event Horizon Telescope, boasting a resolution of $\theta=10^{-5}$ arc-seconds, has successfully imaged structures near the horizon, although its current operations are conducted in the radio band.
	As depicted in the aforementioned \ref{resolution_vs_beta}, the required resolution for observing the vicinity of the black hole's horizon decreases in tandem with the increased value of the extra-dimensional parameter($\beta$) in negative direction. If observations at a lower resolution manage to match the predicted diminished luminosity, it would provide substantial support for the existence of extra dimensions within the theoretical framework. Therefore, it becomes imperative to explore the X-ray to gamma-ray frequency range $\approx$($10^2-10^6$ eV) and to achieve a resolution for observing the photon sphere in the ballpark of $\theta\lesssim 10^{-5}$ arc-seconds for a Schwarzschild black hole, and the corresponding resolution values for observing brane-world black holes, as indicated in the figure above.
	
	\section{Conclusion}\label{Sec:Conclusion}
	If axions are present in the universe, photons traveling through a magnetic field can convert into axions through the coupling term $g_{\Phi\gamma}\Phi F^{\mu\nu}\tilde{F}^{\mu\nu}$. This paper focuses on investigating the phenomenon of photon-axion conversion around black holes. By considering a magnetic field of approximately $\abs{B} \sim (1-30)$ gauss and an axion-photon coupling $\sim10^{-11} \mathrm{GeV}^{-1}$, electron number density $10^4 \mathrm{cm}^{-3}$ we find that the propagation length required for conversion is of the order of a milli-parsec, comparable to the Schwarzschild radius of a supermassive black hole with a mass of $10^9 M_{\odot}$.
	
	Our studies also demonstrate that photons in the X-ray and gamma-ray bands can efficiently convert into axions for masses $\lesssim$ 100 neV. Thus, observing the vicinity of the black hole with electromagnetic waves may reveal a dimming of the photon ring in those wavelengths, provided sufficiently high resolution is achieved in the future. Larger values of $M$ correspond to greater dimming, making supermassive black holes like M87* excellent candidates for observing photon ring dimming. While the Event Horizon Telescope has succeeded in imaging the region in the radio band, achieving such high-resolution observations in the X-ray and gamma-ray bands remains a challenge.
	
	In this paper, we have obtained the path length of photons following the geodesic of a spherically symmetric black hole, enabling us to calculate the photon-axion conversion probability in the spacetime of a black hole with spherical symmetry, given the mass of the black hole is known. The general photon-axion conversion formula of \ref{Sec:PH_AX_CONV} is applicable to any spherically symmetric black hole, including the Schwarzschild black hole as a special case which has been derived in \ref{Eq:Axion_Num}. Through our generalization, we have expanded upon the concepts introduced in \cite{Nomura:2022zyy}, resulting in a more inclusive approach. Importantly, we have also enhanced the accuracy by reducing several approximations in our calculations. Furthermore, by numerically solving the spectra, we have bolstered the credibility of the conversion factor for the Schwarzschild black hole case. 
	
	From our analysis, we have found that in presence of extra dimension for supermassive black holes, the maximum conversion possible is 25\%, and the range of axion mass that exhibits a finite conversion factor lies approximately between (1 - 100) neV. Observations related to the dimming of the photon ring demand very high resolution, approaching $\theta \sim 10^{-5}$ arcsec if the black hole is a Schwarzschild black hole. Similarly, for the black holes with positive $\beta$ values also require resolution at that order of magnitude, as shown in \ref{resolution_vs_beta}. On the other hand, the brane world black hole(i.e for negative $\beta$ values) case would require a lower resolution with a higher charge parameter.      
	In \ref{Fading of the ring}, we explored an approximate estimation of the luminosity after conversion and its related flux, considering the mass of M87*, nearby plasma density, and axion properties such as coupling and mass. If this estimation proves to be accurate in the future, it could offer valuable insights into the coupling and mass of axions. 
	
	This study is based on calculations that consider a uniform magnetic field surrounding the photon sphere. However, it is possible to conduct an analysis considering an inhomogeneous magnetic field. The plasma near the photon sphere is assumed to be homogeneous, but in realistic scenarios, some level of inhomogeneity might be necessary. In \ref{Sec:PH_AX_CONV}, the formula for the total probability of photon-axion conversion was derived in Minkowskian spacetime. The rationale behind using a flat metric for the probability formula, even in the curved geometry of a black hole, is due to our focus on the geodesic motion of photons, which can be accurately represented within a local coordinate system. However, a more comprehensive formula that considers the curved spacetime effects can also be rigorously formulated\cite{Capolupo:2019xyd}. The observations give insight on possessing a spin about the M87*\cite{EventHorizonTelescope:2019pgp,Kawashima:2019ljv}. However, for simplicity, we do not take the rotation of the black hole in the present work. The calculations for a rotating case i.e. a Kerr space time\cite{chandrasekhar1992mathematical,PhysRevD.101.044032} will be more realistic and robust and we will keep it for a future prospect. Even a rotating and charged black hole i.e Kerr-Sen black hole can also be pursued. The polarization of photons which are escaping the photon sphere can be affected by the conversion mechanism\cite{Dessert:2022yqq}. So analysis of these polarized emissions would give valuable insights into axion properties. It is also worth studying the photon axion conversion in the presence of cosmic axion background\cite{Dror:2021nyr,Kar:2022ngx}. The dark matter aspects can also be explored if this axion background is present near the photon sphere. The fading or brightening of the photon sphere produced by black hole mergers can also be modified by the presence of photon axion conversion mechanisms\cite{Arvanitaki:2009fg,Brito:2015oca,Day:2019bbh}. To have a higher flux of axions we can explore the same mechanism for even heavier black holes or a collection of black holes. The cosmological expansion and relative velocity of the galaxy are not taken into account in the present work but for galactic black holes which are even farther away we should have to consider this. The sources which emit the photons through a radiative bremsstrahlung process are considered to be present outside the photon sphere. As the luminosity depends on the lower limit of the integral \ref{Eq:PhNum_NC} than the upper limit more dominantly, it is more important which distance we should take from the black hole centre. In the present work, we have taken the lower limit to be the innermost stable circular orbit or ISCO.	
	\section{Acknowledgement}
	PS acknowledges Tanmoy Kumar, Shauvik Biswas and Vikramaditya Mondal for insightful discussions and useful suggestions. PS is supported by  University Grants Commission, Government of India under the Junior Research Fellowship Scheme. SS is thankful to Sougata Ganguly for helpful discussions while preparing the plots.

	\bibliographystyle{jhep}
	\bibliography{References_2}

\end{document}